\documentclass[aps,pre,eqsecnum,showpacs,draft]{revtex4} 
\usepackage{amsmath,bm}

\begin{document}
 
%\documentstyle[12pt,aps,pre,psfig]{revtex} 
%\documentstyle[preprint,aps,pre]{revtex} %\begin{document}
%\draft 
\title{Instanton versus traditional WKB approach to Landau - Zener problem.}
 
\author{V.A.Benderskii} 
\affiliation {Institute of Problems of Chemical Physics, RAS \\ 142432 Moscow
Region, Chernogolovka, Russia} 
\affiliation{Laue-Langevin Institute, F-38042,
Grenoble, France} 
 
\author{E.V.Vetoshkin} 
\affiliation {Institute of Problems of Chemical Physics, RAS \\ 142432 Moscow
Region, Chernogolovka, Russia} 
\author{E. I. Kats} \affiliation{Laue-Langevin Institute, F-38042,
Grenoble, France} 
\affiliation{L. D. Landau Institute for Theoretical Physics, RAS, Moscow, Russia}
 
\date{\today}

\begin{abstract}
Different theoretical approaches to the famous two state
Landau - Zener problem are briefly discussed.
Apart from traditional methods of the adiabatic perturbation
theory, Born - Oppenheimer approximation with geometric
phase effects, two-level approach, and momentum space representation,
the problem is treated semiclassically also in the coordinate space.
Within the framework of the instanton approach we present
a full and unified description of $1D$ Landau-Zener problem
of level crossing. The method enables us to treat accurately
all four transition points (appearing at two levels crossing),
while the standard WKB approach takes into account only
two of them. The latter approximation is adequate for
calculating of the transition probability or for studying
of scattering processes, however it does not work for finding
corresponding chemical reactions rates, where very often
for typical range of parameters all four transition points
can be relevant.
Applications of the method and of the results may concern
the various systems in physics, chemistry and biology.

\end{abstract}

\pacs{PACS numbers: 05.45.-a, 72.10.-d}
\maketitle
\section{Introduction}
At first sight, the title of this paper might sound perplexing.
What else can be said about Landau - Zener (LZ) problem after numerous
descriptions in both research and textbook literature?
However, although theoretical (and experimental) investigations of different 
LZ systems began
more than seventy years ago, it is still remains an active area of research.
Various approaches to LZ problem have appeared separately in the literature
(see e.g. by no means not a complete list of publications \cite{LL65} - \cite{SK02}), 
these references are not
fully consistent with
each other. We, therefore, think that it is important to collect and discuss
all these approaches in one place.
We will study the $1D$ LZ problem \cite{LL65} of quantum mechanical
transitions between the levels of a two-level system at the avoided
level crossing. 
LZ theory treats a quantum system placed in a a slowly varying external field.
Naturally in such a condition the system adiabatically follows variation
of an initially prepared discrete state until its time dependent energy
level crosses another one. Near the crossing point evidently
the adiabaticity condition is violated (like a semiclassical
behavior is violated near turning points). The slow variation of
the perturbation means that the duration of the transition
process is very long, and therefore the change in the action
during this time is large. In this sense the LZ problem is a semiclassical
one (but with respect to time instead of a coordinate for standard
semiclassical problems).

As is well known the problem presents the most basic
model of non-adiabatic transition which plays very important role in many fields
of physics, chemistry, and biology. Not surprisingly therefore that numerous
monographs and uncountable papers have been devoted to this subject.
In the literature there are roughly speaking three kinds of semiclassical
modelling of LZ problem, namely:
\begin{itemize}
\item
(i) two level system approach \cite{NI62} - \cite{GG01};
\item
(ii)
adiabatic perturbation theory \cite{DY60} - \cite{VS99} (see also
review paper \cite{CH78}) ;
\item
(iii)
momentum space representation \cite{DT72} - \cite{ZN94} .
\end{itemize}
Because different approaches have been proposed to study the LZ problem
one of the immediate motivation of the present paper is to develop an
uniform and systematic procedure for handling the problem.
We will show that all three methods are equivalent
for treating tunneling and over-barrier regions of parameters,
and no one of them can be applied to study say intermediate
region of parameters, where all four of the involving into LZ
system states are relevant. To study this region is our main concern
in this paper.
A second question addressed here concerns so-called connection matrices.
At usual textbook treatment of the LZ problem, only the
transition probabilities are calculated and expressed in terms of
the genuine two-level LZ formula successively applied at each diabatic level
intersection. Evidently such a procedure is an approximation for
the general LZ problem including even in the simplest case 
at least 4 energy levels.
To solve many important physical or chemical problems one must find the $4 \times 4$
(not only $2 \times 2$ connection matrices related these 4 states.

While our paper is not intended as a comprehensive review we detail
here key results of the standard WKB and instanton approaches 
from our own researches and literatures
within the context of different factors that we feel are important
to studying LZ problem. Specifically we focus in the next section (\ref{II})
on the Born - Oppenheimer approximation which is a benchmark
to test semiclassical approximations. In section (\ref{III})
we lay the foundation for treating LZ problem, namely
- adiabatic perturbation theory. Section (\ref{IV}) is devoted 
to
the generalization of the instanton method enables us to investigate
LZ problem in the momentum space. 
It is shown in this section that for a linear (in a $1D$ coordinate
under consideration) potential WKB semiclassical wave functions
in the momentum space coincide with the instanton wave functions.
For the quadratically approximated (parabolic) potentials
the instanton wave functions are exact and have no singularities
(unlike WKB wave functions; remind that the same kind of
relations hold for the WKB and instanton wave
functions in the coordinate space \cite{BM94} - \cite{BV02}).

In this paper we are advocating for the instanton approach,
but it is worth noting that, nevertheless, many important results have been
obtained in the
frame work of the
WKB approach \cite{LL65} -\cite{GG01}. 
For example, one of the very efficient technique (so-called propagator
method) was proposed and elaborated by Miller and his coworkers \cite{MG72},
\cite{PS74}, \cite{MG75} (see also \cite{BM94}). This approach
uses semiclassic (van Fleck - Gutzwiller types) propagators, taking
into account automatically in terms of the general WKB formalism,
the contribution coming from the contour around a complex turning point.
The accuracy of the WKB method can be  
improved considerably, \cite{PK61} \cite{NI62}, \cite{BN65}, \cite{NI68}
(more recent references on so-called Laplace contour integration can be found
also in \cite{AS92})
by the appropriate choice of the integration path around
the turning point, and it appears to be quite accurate
for the tunneling and over-barrier regions, but it becomes non-adequate
in the intermediate energy region.
It has been overlooked in the previous investigations treating
this region by a simple interpolation
from the tunneling (with monotonic decay of the transition probability)
to the over-barrier (with oscillating behavior) regions.

We present all details of the LZ problem for two electronic states in
section \ref{V} using
the instanton description of LZ problem in the coordinate space.
In the section the basic two second order differential (Schr\"odinger) equations
to be dealt with are written in the so-called diabatic state representation
(i.e. in the basis of ''crossed'' levels). Neglecting higher order space
derivatives we find asymptotic solutions, and using adiabatic - diabatic
transformation we match the solutions in the intermediate region.
The complete scattering matrix for the LZ problem is derived in 
the section \ref{VI}.
In the section \ref{VII} we
derive the quantization rules for crossing diabatic potentials
and discuss shortly the application of the obtained
results to some particular models of level crossings 
which are relevant for the interpretation and description of
experimental data on spectroscopy of non-rigid molecules,
on inelastic atomic collisions \cite{NU84}, non-radiative
transitions arising from
''intersystem'' crossings of potential energy surfaces in molecular
spectroscopy and chemical dynamics 
(see e.g. \cite{BM94} and references
herein). 
In the last section \ref{VIII} we draw our conclusions.

Throughout what follows we will consider $1D$ case only. 
The LZ problem for 1D potentials coupled with the thermal
bath of harmonic oscillators is shown to reduced to a certain 
renormalization of the Massey parameter, where entering the
expression for this parameter longitudinal velocity is
decreased due to coupling to transverse 
oscillations (see \cite{BM94} and references herein, and for more
recent references also \cite{KN99}, \cite{SK02}).
Of course the energetic
profile of any real system is characterized by a multidimensional
surface. However, it is often possible to identify a reaction coordinate,
such that the energy barrier between initial
and final states is minimized along this specific
direction, and, therefore,
effectively one can treat the system under consideration
as $1D$. In certain systems, the physical
interpretation of the reaction coordinate is immediate
(e.g. the relative bond length in two diatomic molecules),
but sometimes it is not an easy (if possible at all) task, because
of the many possibilities involved. The latter (multidimensional) case
will be studied elsewhere.
Unfortunately the accuracy of the WKB method near
the barrier top is very poor to make any numbers realistic 
and it is one more motivation to use the alternative to
WKB semiclassical formalism - extreme tunneling trajectory or
instanton technique.

\section{Born - Oppenheimer approximation.}
\label{II}
It may be useful to illustrate the essential physics of the LZ problem
starting with a very well known picture corresponding to the Born - Oppenheimer
approximation \cite{SL63}, \cite{LL65} which leads to the separation
of nuclear and electronic motions, and the approximation is valid only
because electrons are so much lighter than nuclei and therefore
move so much faster.
Thus the small parameter of the Born - Oppenheimer approximation is
\begin{eqnarray}
\label{II1}
\lambda = \left (\frac{m_e}{m}\right )^{1/4} \ll 1 \, ,
\end{eqnarray}
where $m_e$ and $m$ are electronic and nuclear masses respectively.
On the other hands the semiclassical parameter
\begin{eqnarray}
\label{II2}
\gamma  = \frac{m \Omega a^2}{\hbar } \gg 1 \, ,
\end{eqnarray}
where $a$ is a characteristic length of the problem, 
and the characteristic nuclear vibration frequency $\Omega \propto m^{-1/2}$,
therefore $\gamma \propto \lambda ^{-2}$.
From this simple fact important conclusions are arrived at.
Indeed one can satisfy the semiclassical condition $\gamma \gg 1$ by assuming
formally $\hbar \to 0$ or equivalently $\lambda \to 0$. 
This correspondence allows us to apply on the same
footing to the separation of scales for nuclear and electronic motions
either the Born - Oppenheimer or the semiclassical approximation.

In the traditional Born - Oppenheimer approach the solution $\Psi $ 
to the full (i.e. including electronic Hamiltonian $H_e$, depending
on electronic coordinates $r$, and nuclear Hamiltonian depending on
nuclear coordinates $R$) Schr\"odinger equation is presented in the form
of an expansion over the electronic Hamiltonian eigen functions
$\phi _n$
\begin{eqnarray}
\label{II3}
\Psi  = \sum _{n} \Phi _n(R) \phi _n(r, R) \, .
\end{eqnarray}
The electronic eigen values $E_n$
depends on the nuclear coordinates, and the expansion coefficients
$\Phi _n(R)$ is determined by the Born - Oppenheimer equations
\begin{eqnarray}
\label{II4}
\left [-\frac{\hbar ^2}{2 m} \nabla ^2_R + E_n(R) +
\frac{\hbar ^2}{2 m} \sum_{k \neq n} A_{n k} A_{k n} - E \right ] \phi _n = 
-\frac{\hbar ^2}{2 m} \sum _{k , m \neq n}(\delta _{n k} \nabla _R - i A_{n k})(\delta _{k m}
\nabla _R - i A_{k m})\phi _m \, ,
\end{eqnarray}
where for $m \neq k$
\begin{eqnarray}
\label{II5}
A_{m k} = i \langle \phi _m|\nabla _R \phi _k\rangle \, ,
\end{eqnarray}
and all the diagonal matrix elements $A_{n n} = 0$.

Thus from (\ref{II4}) we can find that in the electronic
eigen state $E_n$, the nuclei are moving in the effective potential
\begin{eqnarray}
\label{II6}
U_n(R) = E_n(R) + \frac{\hbar ^2}{2 m} \sum_{k \neq n} A_{n k} A_{k n} \, ,
\end{eqnarray}
and transitions between the electronic states $n$ and $m$ are related
to the non-adiabatic operator in the r.h.s. of (\ref{II4}).
This simple observation allows us to rewrite the effective potential
(\ref{II6}) as
\begin{eqnarray}
\label{II7}
U_n(R) = E_n(R) -\frac{\hbar ^2}{2 m}\sum_{m \neq n}\frac{\langle \phi _n|\nabla _R H_e|\phi
_m\rangle \langle \phi _m|\nabla _R H_e|\phi _n\rangle }{(E_n - E_m)^2}  \, ,
\end{eqnarray}
and from this seemingly trivial expression we arrive at the following
important conclusions:

(i)corrections to $E_n$ have the same order $O(\gamma ^{-2})$ as the
ratio of the nuclear kinetic energy to the potential;

(ii) off-diagonal matrix elements of the non-adiabatic perturbation operator
are also small ($\propto O(\gamma ^{-2})$), and the fact is formulated as the so-called
adiabatic theorem stating that at the adiabatic perturbations ($\lambda \to 0$)
there are no transitions between unperturbed states.

Since the non-adiabatic effects are characterized by the only small parameter
$\gamma ^{-1}$ (the semiclassical parameter), the 
effects can be described in the frame work of semiclassical approaches
(e.g. WKB or instanton ones).
However, one has to bear in mind the main drawback problem
of the Born - Oppenheimer method. Indeed the approximation 
assumes that the electronic wave functions
are real valued ones and formed the complete basis, but it
is impossible to construct such a basis in the whole space
including classically accessible and forbidden regions.

If one relaxes the requirement to have a real valued basis, the 
diagonal matrix elements $A_{nn}
\neq 0$, and the effective adiabatic part of the Born - Oppenheimer
Hamiltonian takes the form
\begin{eqnarray}
\label{II8}
\hat H_n = U_n(R)  + \frac{\hbar ^2}{2 m}(\nabla _R  - i A_{nn}(R))^2
\end{eqnarray}
analogously to the Hamiltonian of a charged particle in a magnetic field
$ B  \propto \nabla _R \times A_{n n}$. Therefore one can change the phases 
of the electronic and nuclear wave functions
\begin{eqnarray}
\label{II9}
\phi _n \to \phi _n \exp (i \chi _n(R)) \, , \,
\Phi _n \to \Phi _n \exp (- i \chi _n (R)) 
\, ,
\end{eqnarray}
by changing respectively the ''vector potential''
\begin{eqnarray}
\label{II10}
A_{nn}(R) \to A_{n n}(R) + \nabla _R \chi _n(R)  \, .
\end{eqnarray}
Thus we confront to an important and, at times, mysterious concept
of the geometrical (or Berry) phase factor that a
quantum mechanical wave function acquires upon a cyclic evolution
\cite{AB59} - \cite{FJ98}.
What is most characteristic for the concept of Berry phase
is the existence of a continuous parameter space in which the state
of the system can travel on a closed path. In our case the phase is determined
by the non-adiabatic interaction (for more details related
to the geometric phase for the Born - Oppenheimer systems, see e.g. the review article
\cite{ME92}).
The phenomenom (which was manifested itself originally as a certain
extra phase shift, appearing upon some external parameter cyclic evolution)
has been generalized for the non-adiabatic, non-cyclic, and non-unitary
cases \cite{AA87}, \cite{SB88}, although the most of the Berry phase
applications concern the systems undergoing the adiabatic evolution
(see e.g. the review article \cite{MS93}).
Note also that apart from the Berry phase some higher order corrections to the
Born - Oppenheimer approximation (traditionally slightly confusing
referred as  geometric magnetism or deterministic friction, see \cite{BR93})
also occur. The practically useful application of the Berry phase conception
is the energy level displacements predicted in \cite{MS86} and observed
by NMR \cite{SC87}.

The essential physics of the phenomena can be illustrated as follows.
There are two subsystems, the fast and the slow ones. The fast
subsystem acquires the Berry phase due to the evolution of the slow subsystem.
In own turn, there is say a feedback effect of the geometric phase
on the slow subsystem. As a result the latter one is framed
by a gauge field affecting its evolution. The gauge field
produces additional forces (Lorentz-like and electric field-like ones)
which have to be included into the classical equation of motion.
In the case of stochastic external forces (e.g. from surrounding
thermal fluctuation media), the Berry phase produces some level broadening for
the fast subsystem. In the limit of low temperatures and strong damping, the slow
subsystem dynamics can be described by the Langevin type equations \cite{GA98}.
The general message we can learn from this fact, is that the geometric phases
are sources of the dissipative processes for LZ systems.

Thanks to its fundamental origin, this geometric phase has attracted
considerable theoretical and experimental attention, however,
its experimentally observable consequences until now have been
scarce. Therefore each opportunity of improving this situation
is worth trying. In this respect the Born - Oppenheimer geometrical
phase provides a unique opportunity for its observation since it
must appear as a non-adiabatic contribution in a standard
Bohr - Sommerfeld quantization rule
\begin{eqnarray}
\label{IIb}
S_n^0 + \chi _n = 2\pi \hbar  \, ,
\end{eqnarray}
where $S_n^0$ is the adiabatic action.

Note that care must be taken when $|E_n(R) - E_m(R)|$ becomes
smaller with respect to the characteristic nuclear oscillation
energy $\hbar \Omega $. It means that in the adiabatic representation
(\ref{II4}) one may not consider the non-adiabatic interaction energy
as a small perturbation. Fortunately in the limit
$$
|E_n(R) - E_m(R)| < \hbar \Omega 
$$
we can start from the other limit with crossing weakly coupled diabatic states,
and consider the adiabatic coupling as a perturbation.
Thus we need the adiabatic - diabatic transformations
enabling us to perform the procedure explicitely, which read
for the wave functions
\begin{eqnarray}
\label{II11}
\tilde \Phi (R) = \exp (i \theta \sigma _y) \Phi (R)  \, ,
\end{eqnarray}
and for the Hamiltonians
\begin{eqnarray}
\label{II12}
\tilde H = \exp (i \theta \sigma _y)H \exp (-i \theta \sigma _y)  \, ,
\end{eqnarray}
where $(H , \Phi )$ and $(\tilde H , \tilde \Phi )$ are the adiabatic and diabatic
representations respectively, $\sigma _y$ is the corresponding Pauli matrix,
and $\theta $ is the parameter of the adiabatic - diabatic transformation
(so-called adiabatic angle).

To illustrate how it works let us consider two coupled ($U_{12}$ is the coupling
energy) crossing effective electronic potentials $U_1(R)$ and $U_2(R)$.
The corresponding adiabatic and diabatic Hamiltonians
are
\begin{eqnarray}
\label{II13} &&
H =  - \frac{\hbar ^2}{2 m}(\nabla _R)^2 + \frac{1}{2}(U_1 + U_2) +
\\ &&
%\end{eqnarray}
%\begin{eqnarray}
\nonumber
\left (\frac{1}{2}(U_1 - U_2)\cos 2\theta (R) + U_{12}\sin 2 \theta (R)\right )\sigma _3
+ \frac{1}{2}\left (-\frac{1}{2}(U_1 - U_2)\sin 2\theta (R) + U_{12} \cos 2\theta (R)\right
)\sigma _1 \, ,
\end{eqnarray}
and
\begin{eqnarray}
\label{II14}
\tilde H = - \frac{\hbar ^2}{2 m}(\nabla _R)^2 + \frac{1}{2}(U_1 + U_2) +
\frac{1}{2}(U_1 - U_2)\sigma _3 U_{12}\sigma _1 \, ,
\end{eqnarray}
where $\sigma _{1 , 2 , 3}$ are the Pauli matrices, and the adiabatic angle
is chosen to eliminate the principle interaction term               
between the adiabatic states
\begin{eqnarray}
\label{II15}
\cot 2 \theta (R)  =  \frac{U_1 - U_2}{2 U_{12}}
\, .
\end{eqnarray}
The adiabatic - diabatic transformation can be brought also into more elegant form
\cite{HP77}, \cite{BL00} 
\begin{eqnarray}
\label{II16}
(\nabla _R - i \hat A)\hat T = 0
\, ,
\end{eqnarray}
where $\hat T$ is the sought for transformation matrix,
and the matrix $\hat A \equiv A_{n n}$ was introduced above (see (\ref{II5})).
The formal solution of Eq. (\ref{II16}) can be represented as a contour
integral
\begin{eqnarray}
\label{II17}
\hat T(s) = \hat T(s_0) \exp \left (- \int _{s_0}^{s} \hat A(s^\prime )d s^\prime \right )
\, ,
\end{eqnarray}
where $s_0$ and $s$ stem for the initial and final points of the contour,
and the solution (\ref{II17}) determines uniquely the transformation
matrix $\hat T$ for the curl-less field $\hat A$
\begin{eqnarray}
\label{II18}
\hat T(t_0) = \hat D \hat T(0)  \, ,
\end{eqnarray}
and the diagonal matrix $\hat D$ can be found from (\ref{II16}) and can be expressed
in terms of the geometric phase factor
\begin{eqnarray}
\label{II19}
D_{k n} = \delta _{k n} \exp (i \chi _k)  \, .
\end{eqnarray}
These two relations (\ref{IIb}) and (\ref{II19}) provide a complete
account of the non-adiabatic transitions, the cornerstone of the LZ problem.
Besides (\ref{IIb}), (\ref{II19}) show that the geometrical Born - Oppenheimer
phases occur from the diabatic potentials crossing points, and enter
quantization rules additively with the contributions from the turning points.
Therefore, our main conclusion from this section is that non-adiabatic
phenomena should be (and can be) included into the general 
semiclassical approach scheme by use of the corresponding connection matrices
\cite{HE62} (see also  \cite{BV02}) for the appropriate combinations
of crossing and turning points of the problem.

\section{Adiabatic perturbation theory.}
\label{III}
It is almost a common student wisdom nowadays that any solution to the adiabatically
time dependent Schr\"odinger equation can be represented
as an expansion over the complete set of stationary
(time independent) eigen functions \cite{LL65}.
For the case under investigation (two level crossing for the
electronic Hamiltonian $H_e(r , t)$), this expansion reads
\begin{eqnarray}
\label{III1}
\Psi (r , t) = c_1(t) \phi _1 (r) + c_2(t) \phi _2(r)  \, ,
\end{eqnarray}
where $\phi _{1 , 2}$ are the stationary with respect to a nuclear motion wave functions.
The time dependent Schr\"odinger equation can be rewritten exactly in the
form of two first order (over time derivatives) equations for $c_1$ and $c_2$
\begin{eqnarray}
\label{III2}
i\hbar
\left (
\begin{array}{c}
\dot c_1 \\ \dot c_2
\end{array}
\right ) =
\left (
\begin{array}{cc}
\tilde H_{11} & \tilde H_{12} \\
\tilde H_{21} & \tilde H_{22} 
\end{array}
\right )
\left (
\begin{array}{c}
c_1 \\
c_2
\end{array}
\right )
\, ,
\end{eqnarray}
where
\begin{eqnarray}
\label{III3}
\tilde H_{k k^\prime } = \langle \phi _k |\tilde H (t)|\phi _{k^\prime }\rangle
\, , \quad  k , k^\prime = 1 , 2
\end{eqnarray}
are the matrix elements for the diabatic Hamiltonian.

The following phase transformation (see, \cite{DY62}, \cite{CH78}, \cite{GG01})
\begin{eqnarray}
\label{III4}
c_k(t) = a_k(t) \exp \left (-\frac{i}{\hbar } \int \tilde H_{kk}(t) dt \right )
\end{eqnarray}
reduces (\ref{III2}) to the coupled first order equations
\begin{eqnarray}
\label{III5}
i \hbar \dot a_1 = \tilde H_{12}a_2 \exp \left (i \int \Omega _{12}(t) dt \right )
\, , \quad
i \hbar \dot a_2 = \tilde H_{21}a_1 \exp \left (- i \int \Omega _{12}(t) dt \right )
\, , 
\end{eqnarray}
where
\begin{eqnarray}
\label{III6}
\Omega _{12} = \frac{1}{\hbar }(\tilde H_{22} - \tilde H_{11})
\, .
\end{eqnarray}
However the slightly different phase transformation
\begin{eqnarray}
\label{III7}
c_k(t) = \tilde \Phi _k(t) \exp \left (\frac{i}{2 \hbar } \int (\tilde H_{11}
+ \tilde H_{22}) dt \right ) 
\end{eqnarray}
keeps the second order Schr\"odinger like equations forms for the
diabatic functions $\tilde \Phi _{1 , 2}$
\begin{eqnarray}
\label{III8}
\hbar ^2 \frac{d^2 \tilde \Phi _1}{d t^2} - \left (\left (\frac{\tilde H_{11} - \tilde H_{22}}{2}
\right )^2 + \tilde H_{12}\tilde H_{21} + \frac{i \hbar }{2}\frac{d}{dt}(\tilde H_{11} - \tilde H_{22})
\right )\tilde \Phi _1 = 0
\, , 
\end{eqnarray}
To clarify the mapping of this time dependent perturbation theory
to the two level crossing problem, and the Born - Oppenheimer
approach described in the previous section \ref{II}, let us consider
the two states Born - Oppenheimer equations in the diabatic representation.
From (\ref{II14}) for one active space coordinate $X$, we have
\begin{eqnarray}
\label{III9}
- \frac{\hbar ^2 }{2 m} \frac{d^2 \tilde \Phi _1}{d X^2} + (\tilde H_{11} - E)\tilde
\Phi _1 = \tilde H_{12}\tilde \Phi _2 = 0
\, , 
\end{eqnarray}
and
\begin{eqnarray}
\label{III10}
- \frac{\hbar ^2 }{2 m} \frac{d^2 \tilde \Phi _2}{d X^2} + (\tilde H_{22} - E)\tilde
\Phi _2 = \tilde H_{21}\tilde \Phi _1 = 0
\, .
\end{eqnarray}
The change of the variables
\begin{eqnarray}
\label{III11}
\tilde \Phi _{1 , 2} = \exp (i k_0 X) c_{1 , 2} \, , \quad k_0^2 = \frac{2 m E}{\hbar ^2}
\end{eqnarray}
transforms the two Born - Oppenheimer equations (\ref{III9}), (\ref{III10})
into the two level crossing equation (\ref{III2}) for the slow
time dependent perturbations if one could neglect the second order derivatives
$(\hbar ^2/2 m)d^2 c_{1 , 2}/d X^2$ and to replace the time derivative by
$v d/d X$ ($v= \sqrt {2E/m}$ is velocity). Obviously we recognize
the standard semiclassical approach in the afore-described procedure.

The same kind of the mapping can be performed also for
the adiabatic amplitudes $C_{1 , 2}(t)$ which are related with the
diabatic amplitudes $c_{1 , 2}(t)$ by the adiabatic - diabatic transformation matrix
depending on the adiabatic angle $\theta $
\begin{eqnarray}
\label{III12}
\left (
\begin{array}{c}
C_1(t) \\ C_2(t)
\end{array}
\right ) =
\left (
\begin{array}{cc}
\cos \theta  & \sin \theta \\
-\sin \theta & \cos \theta
\end{array}
\right )
\left (
\begin{array}{c}
c_1(t) \\
c_2(t)
\end{array}
\right )
\, .
\end{eqnarray}
The corresponding to (\ref{III2}) set of the first order equations in the
adiabatic basis
\begin{eqnarray}
\label{III13}
\left (
\begin{array}{c}
\dot C_1 \\ \dot C_2
\end{array}
\right ) =
\left (
\begin{array}{cc}
H_{11} & - i \dot \theta \\
i\dot \theta & H_{22}
\end{array}
\right )
\left (
\begin{array}{c}
C_1 \\
C_2
\end{array}
\right )
\, ,
\end{eqnarray}
where the non-adiabatic coupling coefficient $\dot \theta $
can be related to the off-diagonal operator $A_{12}$ (\ref{II5})
(or to the geometrical phase, see the previous section \ref{II})
\begin{eqnarray}
\label{III14}
i\dot \theta = A_{12} \equiv i \langle \phi _1 |\dot \phi _2\rangle 
\, .
\end{eqnarray}
The transformation (\ref{III11}) enables to reduce the Born - Oppenheimer
equations (for the nuclear wave functions $\Phi _{1 , 2}$ in the adiabatic
representation) to (\ref{III13}) if and only if:
(i) in the spirit of the semiclassical approach to neglect the second order
derivatives ;
(ii) to keep only $\propto k_0$ terms in the non-adiabatic
matrix elements (i.e. neglecting higher order over $1/k_0$ contributions).
Expressions (\ref{III12}) - (\ref{III14}) do allow an entry point
into the adiabatic perturbation theory developed by Landau \cite{LL65}
and Dykhne \cite{DY62}, \cite{DC63} (see also \cite{DP76}, \cite{HP77}).
We will closely follow the same method.

One can make one step further 
and to find the combination of the two level system amplitudes
$a_{1 , 2}$ (\ref{III4}), (\ref{III5})
\begin{eqnarray}
\label{III15}
Y(t) = \Omega _{12}^{-1/2}\exp \left (- \frac{i}{2}\int \Omega _{12} dt \right )a_1
+ i\Omega _{12}^{-1/2}\exp \left ( \frac{i}{2}\int \Omega _{12} dt\right ) a_2
\, ,
\end{eqnarray}
satisfying to the simple equation
\begin{eqnarray}
\label{III16}
\ddot Y(t) + \frac{\Omega _{12}^2}{4} Y = 0
\, ,
\end{eqnarray}
identical to (\ref{III8}), and the both describe oscillations around the crossing point
in the adiabatic potential (inverted adiabatic barrier). Therefore
formally the adiabatic perturbation theory reduces
the level crossing problem to the well known quantum mechanical
phenomenom - over-barrier reflection. Moreover for the latter problem
the reflection coefficient in a full agreement with the adiabatic theorem
is 1.

Since in $1D$ case evidently two adiabatic potentials have no
reel crossing points, the crossing is possible only at complex
values $X$ or $t$
\begin{eqnarray}
\label{III17}
\Omega _{12} (\tau _c) = 0 \, ; \quad U_1 - U_2 = \pm i U_{12}|_{t = \tau _c}
\, .
\end{eqnarray}
In the vicinity of these points from (\ref{III6})
\begin{eqnarray}
\label{III18}
\Omega _{12}  \propto  (t - \tau _c)^{1/2}
\, ,
\end{eqnarray}
therefore
\begin{eqnarray}
\label{III19}
\int \Omega _{12} d t \simeq \frac{2}{3}(t - \tau _c)^{3/2}
\, ,
\end{eqnarray}
i.e. the crossing points
are square root bifurcation points for the function $\Omega _{12}(t)$.
Using Exp. (\ref{III19}) we depicted in the Fig. 1 the Stokes
and anti-Stokes lines for the equation (\ref{III16}).
The diagram shown in this figure is identical to that corresponding
to the treated semiclassical over-barrier reflection problem
with the linear turning points. The transition probability 
$P_{12}$ in the main
approximation is determined
by the integration over the contour $C(\tau _c)$ going around
the bifurcation point $\tau _c$
\begin{eqnarray}
\label{III20}
P_{12} \simeq \exp \left (\frac{2}{\hbar }\oint _{C(\tau _c)} (H_{11} - H_{22}) d t \right )
\, .
\end{eqnarray}

In the simplest form of the LZ problem the diabatic potentials
are assumed as the linear functions of $t$ or (what is
the same because $t = X/v$) $X$ (see Fig. 2 for the illustration)
\begin{eqnarray}
\label{III21}
U_{1(2)} = U^\# \pm F X
\, .
\end{eqnarray}
Putting (\ref{III21}) into the general expression for the transition
probability (\ref{III20}) we find for this case
\begin{eqnarray}
\label{III22}
P_{12} \simeq \exp (- 2 \pi \nu )
\, ,
\end{eqnarray}
where $\nu = U_{12}^2/2\hbar v F$ is so-called Massey parameter, and the velocity
$v = \sqrt {2|E - U^\# |/m}$.

Some comments about the range of the validity of the approximation seem in order.
A question of primary importance for the LZ problem is related to
the semiclassical nature of the phenomenom. To illustrate it let us note
that for $\Omega _{12}^2 = U_{12}^2 + v^2 F^2 X^2$, the Eq. (\ref{III16})
is the Weber equation with respect to the reel point $X=0$ (diabatic potentials
crossing point).
Evidently, this correspondence of two complex conjugated linear
crossing points $\pm \tau _c$ and one reel $X=0$ crossing point for the Weber
equation is the same as that (in the standard semiclassical
treatment of the Schr\"odinger equation) for two linear and one second order
turning points. Thus by the same manner as for any semiclassical problem,
one can apply to LZ problem WKB or instanton methods.
Let us compare the accuracy of the both approaches. If $|E - U^\# | \gg \hbar \Omega $
(remind that $\Omega $ is a characteristic frequency of the adiabatic potentials),
the WKB method considering for this problem two
isolated linear turning points works quite well (it is the limit of $k_0 a\gg 1$,
corresponding to the adiabatic approximation). If it is not
the case we have to use the diabatic representation.

\section{Instanton method in momentum space.}
\label{IV}
This is not the place to explain the instanton method in details.
However, in a stripped down version the approach can be viewed as follows
(see \cite{BM94} - \cite{BV02}, \cite{PO77}, \cite{CO85}).
The recipe to find the instanton is based on the minimization
of the functional of classical action in the space of paths connecting the minima
in the upside-down potential.
As it is well known \cite{LL65} the expansion of an arbitrary
wave function $\Psi (x)$ in terms of the momentum eigenfunctions
is simply a Fourier integral
\begin{eqnarray}
\label{a1}
\Psi (x) = \frac{1}{2\pi \hbar }\int _{-\infty }^{+\infty }
\exp (i p x/\hbar ) \Phi (p) d p
\, .
\end{eqnarray}
The wave function in the momentum representation $\Phi (p)$ in own turn
can be written down in the semiclassical form
\begin{eqnarray}
\label{a2}
\Phi (p) = A(p) \exp (-i W(p)/\hbar ) 
\, ,
\end{eqnarray}
where the action $W(p)$ is determined by a classical trajectory
$x_0(p)$ according to the definition
\begin{eqnarray}
\label{a3}
\frac{d W}{d p} = x_0(p)
\, .
\end{eqnarray}
We use
dimensionless variables: for the energy $\epsilon = E/\Omega _0$, for the potential $V = U/\gamma \Omega
_0$, and for the coordinate $X= x/a_0$, where $E$ and $U$ are corresponding dimensional values for the
energy and for the potential, 
$a_0$ is a characteristic length of the problem, 
e.g. the tunneling distance, $\Omega _0$
is a characteristic frequency, e.g. the oscillation frequency around the potential minimum.
The dimensionless momentum can be defined as
\begin{eqnarray}
\label{a4}
P = \frac{pa_0}{\gamma \hbar}
\, ,
\end{eqnarray}
where $\gamma $ is semiclassical parameter
(remind that $\gamma \equiv m \Omega _0 a_0^2/\hbar $, where $m$ is a mass
of a particle, and we believe that $\gamma \gg 1$).

Introducing the semiclassical form of the wave function in momentum representation
(\ref{a2}) into the standard one particle $1D$ Schr\"odinger equation, one can
transform it into the form
\begin{eqnarray}
\label{a5}
\left [P^2 + 2 {\hat V} \left (X_0 + i \frac{1}{\gamma }\frac{d}{d P}\right )
- \frac{2}{\gamma }\epsilon \right ] A(P)  = 0
\, .
\end{eqnarray}
In the momentum space $\hat V$ is a potential energy operator and it
can be expanded into a semiclassical series over $1/\gamma $ (or what is the
same, over $\hbar $, and further we
will set everywhere $\hbar = 1$, measuring energies in the units of frequency,
except at some intermediate equations where it is necessary for understanding).                                    
The expansion allows us to consider $\hat V$ as a function $V$ of
two independent variables $X_0$ and $d/d P$, and one can get finally
\begin{eqnarray} &&
\label{a6}
V\left (X_0 + \frac{i}{\gamma }\frac{d}{d P}\right ) = 
V(X_0) +  \frac{i}{\gamma }\left (\frac{d V}{d X_0}\frac{d}{d P}
+ \frac{1}{2}\frac {d^2 V}{d X_0^2} \frac{d X_0}{d P} \right )
+ \left (\frac{i}{\gamma }\right )^2 \Biggl [\frac{d^2 V}{d X_0^2}\frac{d^2}{d P^2}
- 
\\ && 
%\end{eqnarray}
%\begin{eqnarray}
\nonumber
\frac{1}{2} \frac{d^3 V}{d X_0^3}\left (\frac{dX_0}{d P} \frac{d }{d P} - 
\frac{1}{3}
\frac{d^2 X_0}{d P^2}\right ) + 
\frac{1}{24}\frac{d^4 V}{d X_0^4}\left (
\frac{d X_0}{d P}\right )^2 \Biggr ] + ....
\, ,
\end{eqnarray}
where the ellipsis represents all higher order expansion terms.

Accordingly to the general semiclassical rules
from (\ref{a5}) and (\ref{a6}) one can easily find that
the first and the second over $\gamma ^{-1}$ order terms
become identically zero,
if the energy dependent trajectory $X_0(P)$ is determined by the equation
\begin{eqnarray}
\label{a7}
P^2 + 2 V(X_0) = \frac{2\epsilon }{\gamma }
\, ,
\end{eqnarray}
and so-called
transport equation (TE) 
\begin{eqnarray}
\label{a8}
\frac{dV}{dX_0}\frac{dA}{d P} + \frac{1}{2} \frac{d^2 V}{d X_0^2}
\frac{d ^2 W}{d P^2}A
\, ,
\end{eqnarray}
is also satisfied.
The solution of TE (\ref{a8}) can be found explicitely, and it reads
\begin{eqnarray}
\label{a9}
A = \left (
\frac{d V}{d X_0}\right )^{-1/2}
\, .
\end{eqnarray}
It follows from (\ref{a9}) that the semiclassical WKB wave function
(\ref{a2}) has singularities in all stationary points
of the potential $V$, i.e. these points are the turning points in the momentum space.
It illustrates 
fundamental difficulties 
of the WKB procedure, i.e. how to match the solutions which
become singular on caustic lines separating manifolds in phase space with real and imaginary momenta.

To illustrate also the second drawback of the WKB method let
us consider the linear ($V = F X$) and harmonic ($V = X^2/2$) potentials.
The trajectories $X_0(P)$ can be trivially determined from (\ref{a7}).
For the linear potential, $X_0(P)$ is the inverted parabola with the maximum
(the top of the inverted parabola) $ X_{0m} = \epsilon F/\gamma $ at $P =0$.
The left and the right branches of the parabola correspond to the opposite
motion directions in the classically accessible region $ X_0 < X_{0m}$.
The semiclassical WKB wave function in the momentum space for the linear potential
\begin{eqnarray}
\label{a99}
\Phi (P) = \frac{1}{\sqrt F} \exp \left (-\frac{i}{F}\left (\epsilon
P - \gamma \frac{P^3}{6}\right ) \right )
\, ,
\end{eqnarray}
is the Fourier transform of the coordinate space Airy function. For the harmonic
potential the corresponding trajectories (\ref{a7}) are the ellipses and the wave
functions in the both spaces (momentum and coordinate) have
the same functional form. It is worthwhile to note 
that although the WKB functions are not exact, the
corresponding 
eigenvalues coincide with the exact quantum mechanical ones.

As we have recently shown \cite{BV99} - \cite{BV02}, one can successfully attack many
important semiclassical problems by the instanton method.
Having in mind in this section momentum space let us remind for the sake of conveniency 
the main ideas of the instanton approach.
The first step of the approach derived in \cite{PO77} and
\cite{CO85} is so-called Wick rotation of
a phase space corresponding to a transformation to imaginary time $ t \to - i t$.
At the transformation, the both, potential and kinetic, energies change their signs,
and Lagrangian is replaced by Hamiltonian in the classical equation
of motion. 
In the momentum space the low energy instanton wave functions
can be constructed using Wick rotation in the momentum space
(i.e. the transformation $P \to i P$) and besides the term with the energy $\epsilon $
in (\ref{a7}) should be removed from this equation and taken into account
in the TE (\ref{a8}). The trajectory $X_0(P)$ in the instanton formalizm
describes zero energy motion in the classically forbidden region of
the momentum space where the wave function has a form
\begin{eqnarray}
\label{a10}
\Phi (P)  = \left (
\frac{d V}{d X_0}\right )^{-1/2} Q(P) \exp (-\gamma W(P))
\, ,
\end{eqnarray}
and additional prefactor $Q(P)$ which has appeared in (\ref{a10}) 
can be represented as
\begin{eqnarray}
\label{a11}
\ln Q(P) = \epsilon \int \left (
\frac{d V}{d X_0}\right )^{-1} dP
\, .
\end{eqnarray}

In the particular case of a linear over $1D$ space coordinate potential
($V(X) = F X$) the instanton and WKB functions have the same form.
For an arbitrary ($n$-order) anharmonic potential, the Schr\"odinger equation
in the momentum space is reduced to the $n$-th order differential
equation. However the derivatives of the $n$-th order decrease
proportional to $\gamma ^{-n}$, and therefore corresponding terms
can be taken into account perturbatively. A rigorous mathematical method
to perform this procedure (we will use in our paper) has been developed
by Fedoryuk \cite{FE64} - \cite{FE66}.

To illustrate the instanton approach let us consider the simplest
form of the LZ problem depicted in the Fig. 3 (see all notation
in the figure caption). For the linear with arbitrary line slopes
potentials in the diabatic state representation one has two second order
coupled equations 
\begin{eqnarray} &&
\label{a12}
- \frac{d^2 \Theta _1}{d X^2} = \gamma ^2 (\alpha + f_1 X) \Theta _1 =
\gamma ^2 v \Theta _2 \, ; \\ &&
%\end{eqnarray}
%\begin{eqnarray}
\nonumber
- \frac{d^2 \Theta _2}{d X^2} = \gamma ^2 (\alpha + f_2 X) \Theta _2 =
\gamma ^2 v \Theta _1
\, ,
\end{eqnarray}
where $\Theta _{1 , 2}$ are the 
eigenfunctions of the corresponding states 
and all other notation are introduced according to the Fig. 3, namely
$$
\Omega ^2 = \frac{a^2 F^2}{m U_{12}} \, , \, F = \sqrt {F_1 |F_2|} \, , \,
\gamma = \frac{a^3 F m^{1/2}}{U_{12}^{1/2}} \, , \,
\alpha = 2\frac{U_0 - E}{\gamma \Omega } \, , \,
f_{1,2} = 2\frac{a F_{1,2}}{\gamma \Omega } \, ,\, v = 2 \frac{U_{12}}{\gamma \Omega } \, .
$$ 
The equations (\ref{a12}) can be transformed into the momentum space
(they formed a coupled linear differential equation system)
and after that can be rewritten as one second order equation
\begin{eqnarray}
\label{a13}
\frac{d^2 \Psi _1}{d k^2} + q(k) \Psi _1 (k) = 0
\, ,
\end{eqnarray}
where we introduced
\begin{eqnarray}
\label{a14}
\Psi _1 = \Phi _1 \exp \left [i\frac{\gamma \alpha ^{3/2}}{2} \left (\frac{1}{f_1}
+ \frac{1}{f_2}\right ) \left (k + \frac{k^3}{3}\right)\right ]
\, ,
\end{eqnarray}
and $\Phi _1$ is the Fourier transformed of $\Theta _1$,
$k = P/\gamma \sqrt \alpha $, and $q(k)$ is a fourth order characteristic polynomial
\begin{eqnarray}
\label{a15}
q(k) = \lambda ^2(1 + k^2)^2 + 2 \lambda (i k - 2 \nu )
\, ,
\end{eqnarray}
dependent of two parameters
\begin{eqnarray}
\label{a16}
\lambda  = \frac{1}{2} \gamma \alpha ^{3/2} \left (\frac{1}{f_1}
- \frac{1}{f_2}\right ) \, ,
\, \nu = \frac{\gamma v^2}{2(f_1 - f_2)\sqrt \alpha } 
\, .
\end{eqnarray}
The first parameter $\lambda $ in the momentum representation
plays the role of the new semiclassical
parameter, the second one is the known 
(and already defined in (\ref{III22})) Massey parameter.

Fortunately all roots of the characteristic polynomial (\ref{a15})
can be found analytically quite accurately in the most interesting
physically region of parameters. To simplify expressions (keeping
complete physical content) we present results only for the simplest
case (symmetric slopes of the diabatic potentials) $f_1 = - f_2 \equiv f$.
In the classically forbidden region $U^\# - E > 0$, $\alpha > 0$, at $\lambda \gg 1$
(i.e. equivalently at $ \alpha \gg (f/\gamma )^{2/3}$), all the four
roots of the polynomial are close to $\pm i$
\begin{eqnarray}
\label{a17}
k_1^\pm = i \left (1 \pm \sqrt {\frac{1 + \nu }{2 \lambda }}\right )
\, , \, k_2^\pm = \pm \sqrt {\frac{1 - \nu }{2 \lambda }} - i
\, .
\end{eqnarray}
In the classically accessible region ($U^\# - E < 0\, , \, \alpha < 0$),
the roots are close to $\pm 1$, if $\lambda \gg 1$ (or if $-\alpha \gg (f/\gamma )^{2/3}$)
\begin{eqnarray}
\label{a18}
k_1^\pm = 1 \pm 
\left (\frac{\sqrt {1 + \tilde \nu ^2} + \tilde \nu}{4 \tilde \lambda }\right )^{1/2}
\pm i \left (\frac{\sqrt {1 + \tilde \nu ^2} - \tilde \nu}{4 \tilde \lambda }\right )^{1/2}\, , \,
k_2^\pm = - 1 \mp
\left (\frac{\sqrt {1 + \tilde \nu ^2} + \tilde \nu}{4 \tilde \lambda }\right )^{1/2}
\pm
\left (\frac{\sqrt {1 + \tilde \nu ^2} - \tilde \nu}{4 \tilde \lambda }\right )^{1/2}
\end{eqnarray}
(the tilde sign means that in the corresponding quantity $\alpha $
should be taken by its modulus, i.e. at $|\alpha |$).

Found above the set of the roots of the characteristic polynomial (\ref{a15})
in the classically forbidden (\ref{a17}) and in the classically accessible (\ref{a18})
regions is formally equivalent to the corresponding transition or turning points
for the system of two potential barriers or two potential wells respectively.
Thus we can use all known for these cases WKB and instanton results (see e.g.
for all details 
our recent publications \cite{BV02} and references herein).
Since in the semiclassical treatment we usually concerned only with the asymptotic
solutions and their connections via transition or turning points on the complex plane,
the famous Stokes phenomenom \cite{HE62}, \cite{PK61} of asymptotic solutions
plays an essential role, and the distribution of the transition points
(which are nothing but the zero points of the characteristic polynomial)
and Stokes and anti-Stokes lines, dictates the phenomenom.
We show all the lines emanating from linear turning points in Fig. 2.
In the case when the roots formed a pair of close linear turning points,
one can replace each such a pair by one second order turning points.
The corresponding Stokes and anti-Stokes lines are depicted in the Fig. 4.

In the classically forbidden region the instanton wave functions
can be found using the roots (\ref{a17})
\begin{eqnarray}
\label{a19}
\Phi ^+_1 =
\frac{(1 - i k)^{\nu - 1}}{(1 + i k)^{\nu + 1}} \exp
\left (i \gamma \left (k + \frac{k^3}{3}\right ) \right )
\, , \,
\Phi ^- _1 = 
\frac{(1 - i k)^{-\nu }}{(1 + i k)^{-\nu }} \exp
\left (-i \gamma \left (k + \frac{k^3}{3}\right ) \right )
\, .
\end{eqnarray}
At $|k| \to \infty $ the function $\Phi ^+_1$ decreases proportional
to $|k|^{-2}$, and $\Phi ^-_1$ is reduced to the Airy function \cite{EM53}, \cite{OL74}.
In the vicinity of the second order turning points $k =\pm i$,
the fourth order characteristic polynomial is reduced to the second order
one, and the equation (\ref{a13}) is reduced to the Weber equation with
the known fundamental solutions \cite{EM53}
$$
D_{-\nu }(\pm 2 \sqrt \lambda (k + i) )\, , \,  
$$
at $|k + i| \to 0$ and
$$  
D_{-\nu - 1} (\pm 2 \sqrt \lambda (k - i) )\, , \,  
$$
at $|k - i| \to 0$.
The same procedure applied to the classically accessible region leads
to the solutions
\begin{eqnarray}
\label{a20}
\Phi ^+_1 =
\frac{(1 - k)^{i \tilde \nu - 1}}{(1 + k)^{i \tilde \nu + 1}} \exp
\left (i \tilde \gamma \left (k - \frac{k^3}{3}\right ) \right )
\, , \,
\Phi ^- _1 = 
\frac{(1 + i k)^{i \tilde \nu }}{(1 + i k)^{i \tilde \nu }} \exp
\left (-i \tilde \gamma \left (k - \frac{k^3}{3}\right ) \right )
\, ,
\end{eqnarray}
and it is reduced to 
the Weber equation fundamental solutions
too
$$
D_{i \tilde \nu }(\pm 2 \sqrt {\tilde \lambda } \exp(i \pi /4)(k + 1) )\, , \,  
$$
at $|k + 1| \to 0$ and
$$  
D_{i \tilde \nu - 1} (\pm 2 \sqrt {\tilde \lambda }\exp(i\pi /4) (k - 1) )\, , \,  
$$
at $|k - 1| \to 0$.

The same solutions can be obtained for LZ problem in the two level approximation
using the instanton method in the coordinate space. The reason for it
is fairly transparent and based on the fact that for linear diabatic
potentials the limit $k \to \pm \infty $ corresponds to the limit
$x \to \pm \infty $, and therefore the asymptotic of the
solutions are the same in the momentum and in the coordinate space.

The whole analysis can be brought into a more compact form by introducing so-called connection
matrices. In the instanton approach we have deal
with the asymptotic solutions and their connections on the complex coordinate plane.
Thus it is important and significant to know the connection matrices. 
The needed connection matrices can be found easily by matching of the solutions
(\ref{a19}) or (\ref{a20}) at the second order turning points through
the corresponding fundamental solutions of the Weber equation.
It gives the following connection matrices
\begin{equation} 
\label{a21} 
\hat M_1 = \left( 
\begin{array}{cc} 
- \cos (\pi \nu ) & \sqrt {2 \pi }\exp(-2\chi )/\Gamma (\nu )   \\ 
\Gamma (\nu )\exp (2 \chi ) \sin ^2(\pi \nu )/\sqrt {2 \pi } & \cos (\pi \nu ) 
\end{array}
\right ) \, ,
\end{equation}
where $ \chi = (\nu - (\nu - 1/2)\ln \nu )/2$, and
\begin{equation} 
\label{a22} 
\hat M_2 = \left( 
\begin{array}{cc} 
- \exp (-\pi \tilde \nu ) & \sqrt {2 \pi }\exp(-\pi \tilde \nu )\exp (-2\tilde \chi )/\Gamma (-i
\tilde \nu )   \\ 
2\Gamma (-i \tilde \nu )\exp (-\pi \tilde \nu /2)\exp (2 \tilde \chi ) sh (\pi \tilde \nu )/\sqrt {2 \pi }
&
\exp (-\pi \tilde \nu ) 
\end{array}
\right ) \, ,
\end{equation}
where $\tilde \chi = \left (-i((\pi /4) + \tilde \nu (1 - \ln (\tilde \nu ))\right ) +
(1/4) \ln \tilde \nu $.

As a note of caution at the end of this section we should
also remind that initially for the linear diabatic potentials
we had two corresponding Schr\"odinger equations, and each of them possesses
two fundamental solutions. Thus the full LZ problem is characterized by
the four fundamental solutions to the left with respect to a given turning point
asymptotic and four fundamental solutions to the right (with respect to
the same turning point) asymptotic. Therefore generally speaking the connection matrices
should be $ 4 \times 4$ ones. However
owing to the symmetry of the potentials these $4 \times 4$ matrices have
two blocks $2 \times 2$ structures for the functions $\Phi _1 $ and $\Phi _2$.
The latter ones are given in (\ref{a21}) and in (\ref{a22}) respectively.

\section{LZ problem for two electron states (instanton approach in the coordinate space).}
\label{V}
In the previous sections \ref{II} - \ref{IV} we investigated 
LZ problem in the frame work of the adiabatic
perturbation theory, two level approximation and momentum
representation. The all three methods are equivalent and semiclassical by their
nature. Correspondingly the approaches do work properly
for the tunneling and over-barrier transmission energy regions,
but become non-adequate within the intermediate region (of the order of $\gamma ^{-2/3}$)
near the level crossing point. 
The fact is that the accuracy of these methods depends on the ''renormalized''
(energy dependent) semiclassical parameter $\lambda $ (\ref{a16}) which can be
small in the intermediate region ($\lambda \leq 1$ even for $\gamma \gg 1$).
To treat this region we have to use the coordinate
space presentation, since we need to know the connection matrices for
non-adiabatic transitions. For the latter problem the wave functions outside
the level crossing point, happen to be more convenient
(and, besides, have a more compact mathematical form) in the
coordinate space.

\subsection{Tunneling and over-barrier regions}
\label{VA}
To move further on smoothly we start here reproducing in the
coordinate space the results found in the previous sections for
the tunneling and over-barrier energy regions. In the diabatic representation
we can rewrite two second order LZ differential equations
(\ref{a12}) as the following fourth order linear differential equation
with the constant coefficients at the derivatives (for the sake of simplicity
we consider the symmetric slope case $ f_1 = - f_2 \equiv f$)
\begin{eqnarray}
\label{b1}
\frac{d^4 \Phi _1}{d X^4} - 2 \gamma ^2 \alpha \frac{d^2 \Phi _1}{d X^2} -
2\gamma ^2 f \frac{d \Phi _1}{d X} + \gamma ^4(\alpha ^2 - v^2 - f^2 X^2)\Phi _1 =0
\, .
\end{eqnarray}
Mathematical formalism elaborated by Fedoryuk \cite{FE64} - \cite{FE66},
we are about to recall, reduces (\ref{b1}) by a semiclassical substitution to
a set of equations of the order $\gamma ^n$. Characteristic polynomial for
(\ref{b1})
\begin{eqnarray}
\label{b2}
F(\lambda ) = \lambda ^4 - 2 \alpha \gamma ^2\lambda ^2 - 2 \gamma ^2 f \lambda
+\gamma ^4(\alpha ^2 - v^2 - f^2 X^2)
\, ,
\end{eqnarray}
where by the definition $\lambda = d W/d X$.

Solving the equation $F(\lambda ) = 0$ perturbatively over $\gamma ^{-1} \ll 1$
we find
\begin{eqnarray}
\label{b3}
\lambda _j = \lambda _j^0 + u_j
\, ,
\end{eqnarray}
where
\begin{eqnarray}
\label{b4}
\lambda _j^0 = \pm \left [\gamma (\alpha \pm \sqrt {v^2 + f^2 X^2})\right ]^{1/2}
\, ,
\end{eqnarray}
and
\begin{eqnarray}
\label{b5}
u_j = \frac{\gamma f}{2}\left [(\lambda _j^0)^2 - \alpha \gamma \right ]^{-1}
\, .
\end{eqnarray}
Four asymptotic solutions of (\ref{b1}) thus can be represented
as
\begin{eqnarray}
\label{b6}
\{ y_j \} \equiv \{ \Phi _+^+ , \Phi _+^- , \Phi _-^+ , \Phi _-^- \} 
= (v^2 + f^2 X^2)^{-1/4}\exp \left [\int
_{0}^X \lambda _j(X^\prime ) d X^\prime \right ]
\, ,
\end{eqnarray}
and these describe a motion with imaginary momentum in the upper and lower 
adiabatic potentials
$$
\frac{2m a^2}{\hbar ^2}\left (U^\pm - E \right ) = \gamma ^2 (\alpha \pm \sqrt {v^2 +
f^2 X^2})
\, .
$$
The subscripts in (\ref{b6}) corresponds to the upper or lower adiabatic levels,
and the superscripts are referred to the sign of the action.

Before turning on the connection matrices let us use
the substitution
\begin{eqnarray}
\label{b7}
\Phi _1 = \exp (\kappa X)\phi 
\, ,
\end{eqnarray}
and choose $\kappa $ value to vanish the first derivative in (\ref{b1}), i.e.
\begin{eqnarray}
\label{b8}
\kappa ^3 - \gamma ^2\alpha \kappa - \frac{1}{2} \gamma ^2 f = 0
\, .
\end{eqnarray}
At $\alpha > 3(f/4\gamma )^{2/3}$ one can expand the roots of (\ref{b8})
in terms of the parameter
\begin{eqnarray}
\label{b9}
\delta = \frac{f}{4\gamma }\alpha ^{-3/2} < \frac{1}{3 \sqrt 3}
\, .
\end{eqnarray}
Thus we find
\begin{eqnarray}
\label{b10}
\kappa _1 = \gamma \sqrt \alpha \left (1 + \frac{\delta }{2}\right )
\, , \, \kappa _2 = \gamma \sqrt \alpha \left (-1 + \frac{\delta }{2}\right )
\, , \kappa _3 = \gamma \sqrt \alpha \delta
\, .
\end{eqnarray}
At the condition (\ref{b8}) the coefficients at the fourth and at the
third order derivatives in (\ref{b1}) are small (proportional to $\delta $
and to $\sqrt \delta $ respectively) and the fourth order equation (\ref{b1})
can be rewritten as two second order Weber equations with the solutions
$$
D_{p^{(1 , 2)}}(\beta _{(1 , 2)} X)
\, ,
$$
where
\begin{eqnarray}
\label{b11}
p^1 = - 1 + \frac{\delta }{2} - \nu \left (1 - \frac{3 \delta }{2} \right )
\, , \, p^2 = \frac{\delta }{2} - \nu \left (1 + \frac{3 \delta }{2} \right )
\, , \beta _{(1 , 2)} = \left (\frac{\gamma ^2 f^2}{\alpha }\right )^{1/4}
\left (1 \pm \frac{3 \delta }{4} \right )
\, .
\end{eqnarray}
The leading terms of these solutions are the same as found in
the previous section \ref{IV}. But the Fedoryuk method we used, gives us also
in the tunneling region (\ref{b8}) the higher order over the parameter
$\delta $ corrections.

In the over-barrier energy region where $\alpha < - 3(f/4 \gamma)^{2/3}$,
the roots of the equation (\ref{b8}) are complex - conjugated
\begin{eqnarray}
\label{b12}
\frac{\kappa _{(1 , 2)}}{\gamma \sqrt \alpha } =
-\frac{\tilde \delta }{2} \pm i \left (1 + \frac{3 \tilde \delta ^2 }{8}\right )
\, ,
\end{eqnarray}
and $\tilde \delta $ plays the role of the small parameter in this region
\begin{eqnarray}
\label{b13}
\tilde \delta = \frac{f}{4 \gamma |\alpha |^{3/2}}
\, .
\end{eqnarray}
Again as above for the tunneling region, the coefficients at the
higher order derivatives are small, and therefore, the function $\phi $
(\ref{b7}) satisfies the Weber equation with the fundamental solutions
$$
D_{\tilde p^{(1 , 2)}}(\tilde \beta _{(1 , 2)} X)
\, ,
$$
where
\begin{eqnarray}
\label{b14} 
\tilde p^1 = - 1 + i\frac{3 \tilde \delta }{2} + i \nu \left (1 + \frac{3 \tilde \delta }{4} \right )
\, , \, \tilde p^2 = i\frac{3\tilde \delta }{2} + i \nu \left (1 - \frac{3\tilde \delta }{4} \right )
\, , \, 
\\ 
\nonumber
\tilde \beta _1  = \exp(i\pi /4) \left (\frac{\gamma ^2 f^2}{|\alpha |}\right )^{1/4}
\, , \, \tilde \beta _2 = \exp(-i 3\pi /4) \left (\frac{\gamma ^2 f^2}{|\alpha |}\right )^{1/4}
\, .
\end{eqnarray}
Like it was for the tunneling region (\ref{b11}), the leading terms of
the expansion (\ref{b14}) coincide with the results found in the previous sections,
but from (\ref{b14}) we are able to compute the corrections to the main terms.

Now we are in the position to find the connection matrices.
To do it for the tunneling region
we have to establish the correspondence between the solutions
of the fourth order differential equation (\ref{b1}) and those for the localized
in the left ($L$) and in the right ($R$) wells states.
In the case $\alpha \gg f|X|$ for the diabatic potentials, the action
can be computed starting from the both wells ($R$ and $L$)
\begin{eqnarray}
\label{b15}
\gamma W^L \simeq \gamma W_0^L + k_0 X + \frac{\beta ^2}{4} X^2 \, ,
\gamma W^R \simeq \gamma W_0^R - k_0 X + \frac{\beta ^2}{4} X^2 
\, ,
\end{eqnarray}
where $k_0 = (2m a^2(U^\# - E)/\hbar ^2)^{1/2} \equiv \gamma \sqrt \alpha $ is
imaginary momentum, and $W_0^{L , R}$ are the actions computed from an
arbitrary distant point in the $L$ or in the $R$ wells, respectively to
the point $X=0$.
From the other hand in the adiabatic potentials $U^\pm = U^\# \pm \sqrt {U_{12}^2
+ a^2 f^2 X^2}$ the corresponding actions can be represented
\begin{eqnarray}
\label{b16}
\gamma W^\pm - \gamma W_0^\pm = k_0X \pm \frac{\beta ^2}{4} X^2 sign (X)
\, .
\end{eqnarray}
Explicitely comparing the semiclassical wave functions in the both representations
(adiabatic and diabatic ones) it is easy to see that the adiabatic
functions in the potential $U^-$ coincide with the diabatic functions
for localized $L$ and $R$ states at $X<0$ and $X>0$ respectively. The adiabatic
functions for the upper potential $U^+$ correspond to the tails of the diabatic
wave functions localized in the opposite wells.
Therefore in the level crossing region the $L/R$ diabatic functions 
are transformed into the $R/L$ functions, and the interaction entangles the diabatic
states with the same sign of $k_0 X$. Thus we have only four non-zero amplitudes
of the following transitions
\begin{eqnarray}
\label{b17}
\langle \Phi ^+_L|\Phi ^-_R\rangle \, , \, 
\langle \Phi ^-_L|\Phi ^+_R\rangle \, , \, 
\langle \Phi ^+_R|\Phi ^-_L\rangle \, , \, 
\langle \Phi ^-_R|\Phi ^+_L\rangle 
\, .
\end{eqnarray}
Recalling that
\begin{eqnarray}
\label{b18}
\gamma W^\pm = \gamma \int \left (\alpha \pm \sqrt {v^2 + f^2 X^2}\right )^{1/2}
\simeq k_0 X \pm \frac{\beta ^2}{4} X^2  \pm \frac{\nu }{2}(1 - \ln \nu )
\, ,
\end{eqnarray}
we come to the conclusion that the quantum solutions (\ref{b11}), valid in the vicinity
of the level crossing point asymptotically, match smoothly increasing and decreasing   
solutions (\ref{b6}), and it leads to the Landau description \cite{LL65} of the level crossing
transitions depicted in the Fig. 5.

Using expressions (\ref{a21}), (\ref{a22}) relating the fundamental solutions
of the Weber equation, we can find the corresponding to (\ref{b17})
$4 \times 4$ connection matrix
\begin{eqnarray} && 
\label{b19}
\left( 
\begin{array}{c}
\Phi _R^+ \\
\Phi _R^- \\
\Phi _L^+ \\
\Phi _L^-
\end{array}
\right ) = \\ &&
%\end{equation}
%\begin{equation}
\nonumber 
\left [
\begin{array}{cccc}
\sqrt {2\pi }\exp(-2 \chi )/\Gamma (\nu ) & 0 & 0 & -\cos (\pi \nu ) \\
0 & \Gamma (\nu )\exp (2\chi )\sin ^2(\pi \nu ) & - \cos (\pi \nu ) & 0 \\
0 & \cos (\pi \nu ) & \sqrt {2\pi }\exp (-2 \chi )/\Gamma (\nu ) & 0 \\
\cos (\pi \nu ) & 0 & 0 & \Gamma (\nu ) \exp (2 \chi ) \sin ^2(\pi \nu )/\sqrt {2\pi }
\end{array}
\right ]
\left (
\begin{array}{c}
\Phi _L^- \\
\Phi _L^+ \\
\Phi _R^- \\
\Phi _R^+
\end{array}
\right ) \, ,
\end{eqnarray}
where as above
$$
\chi = \frac{\nu }{2} - \frac{1}{2}\left (\nu - \frac{1}{2} \right ) \ln \nu
\, .
$$
The matrix (\ref{b19}) has a $2 \times 2$ block structure, and each of the
identical blocks connects increasing and decreasing diabatic solutions.
However these diagonal blocks do not correspond to the $L - R$
transitions separately for the lower and upper adiabatic potentials.
Indeed the corresponding to these transitions $2 \times 2$ matrix is
\begin{equation} 
\label{b20}
\left( 
\begin{array}{c}
\Phi _R^+ \\
\Phi _L^-
\end{array}
\right ) =
\left [
\begin{array}{cc}
\sqrt {2\pi }\exp(-2 \chi )/\Gamma (\nu ) & -\cos (\pi \nu ) \\
\cos (\pi \nu )  & \Gamma (\nu )\exp (2\chi )\sin ^2(\pi \nu )/\sqrt {2\pi }
\end{array}
\right ]
\left (
\begin{array}{c}
\Phi _L^- \\
\Phi _R^+
\end{array}
\right ) \, ,
\end{equation}
In the diabatic limit (i.e. $\nu \to 0$) the diagonal matrix elements
are small ($\propto \nu ^{1/2}$, and $\nu ^{3/2}$ respectively),
and the off-diagonal elements approach to $\pm 1$, as it should be since
by the definition there are no transitions between the diabatic potentials.

In the adiabatic limit $\nu \gg 1$, the diagonal matrix elements
tend to 1, and it means that the decreasing $L$ solution transits only
into the increasing $R$ solution, and vice versa. Thus the connection matrix in
the tunneling region depends only on the Massey parameter $\nu $.
One has to bear in mind here that the blocks of the $4 \times 4$
connection matrix (\ref{b19}) correspond to the two isolated 
second order turning points with the Stokes constant
(see e.g. \cite{BV02})
\begin{eqnarray}
\label{b201}
T_2 = \frac{\sqrt {2\pi }}{\Gamma (\nu )} \exp (-2\chi )
\, .
\end{eqnarray}

Analogously one can study the over-barrier region. Repeating again
the procedure described above for the tunneling region (with
evident replacements $k_0 \to -i k_0$ and $\beta ^2 \to i \beta ^2$) we
end up with the following $4 \times 4$ connection matrix
\begin{eqnarray} 
\label{b21} &&
\hat U
= 
\left [
\begin{array}{cc}
\sqrt {2\pi }\exp(-2 \tilde \chi )/\Gamma (-i\nu ) & 0  \\
0 & 2\Gamma (-i\nu )\exp (-\pi \nu )\exp (2\tilde \chi )\sinh (\pi \nu )/\sqrt {2\pi } \\
0 & \exp (-\pi \nu ) \\
\exp (-\pi \nu ) & 0 
\end{array}
\right . \\ &&
%\end{equation}
%\begin{equation} 
\nonumber
\left .
\begin{array}{cc}
 0 & -\exp (-\pi \nu ) \\
- \exp (-\pi \nu
) & 0 \\
\sqrt {2\pi }\exp (-2 \tilde \chi )/\Gamma (-i\nu ) & 0 \\
0 & 2\Gamma (-i\nu ) \exp (2 \tilde \chi ) \exp (-\pi \nu )\sinh (\pi \nu )/\sqrt
{2\pi }
\end{array}
\right ]
\, ,
\end{eqnarray}
where
\begin{eqnarray}
\label{bd}
\tilde \chi = - \frac{i}{2}\left (\frac{\pi }{4} + \nu (1 - \ln \nu)\right )
+ \frac{1}{4}(\pi \nu + \ln \nu )
\, .
\end{eqnarray}
The same manner (as it was already mentioned for the tunneling
region), the blocks in (\ref{b21}) correspond to the two isolated
second order turning points with the Stokes constant
\cite{BV02}
\begin{eqnarray}
\label{b211}
\tilde T_2 = \frac{\sqrt {2\pi }}{\Gamma (-i\nu )} \exp (-2\tilde \chi )
\, .
\end{eqnarray}
Thus we arrive at the important conclusion that the main peculiarity
of the LZ level crossing (in comparison with the standard say one-potential
problems) is that the second order turning points characterizing the
diabatic levels crossing for the LZ problem, possesses the different Stokes constants 
$T_2$ (\ref{b201}) and $\tilde T_2$ (\ref{b211}) in the tunneling and
in the over-barrier regions.

\subsection{Intermediate energy region}
\label{VB}

We can now reap the fruits of the previous subsection consideration
efforts. First, let us note that from the relations (\ref{b11}) and
(\ref{b14}) one can see that when the energy approaches
to the top of the barrier, the exponents $p^{(i)}$ and $\tilde p^{(i)}$
of the parabolic cylinder functions are increased and 
thus, more and more deviated from the value prescribed by the Massey parameter
$\nu $.
Second, increasing of $\beta _{(i)}$ upon $|\alpha |$ decreasing, decreases 
the values of $|X|$ where the asymptotic smooth matching of the solutions
should be performed. For $\delta \to 0$ these $|X|$ values are located deeply in the
classically forbidden region, where the potentials are close to
the diabatic potentials, while for $\delta \geq 2\sqrt 3/3$, these coordinates
$|X|$ are of the order of the quantum zero-point oscillation amplitudes,
and therefore to find the solution in this region, we have to use the adiabatic 
representation.

These two simple observations give us a conjecture how to treat LZ problem 
in the intermediate energy region.
To do it first of all we should find the energetical ''window'' for
the intermediate region. It is convenient to chose the adiabatic potential
frequency $\Omega = F/\sqrt {m U_{12}}$ as the energy scale, and in terms
of this scale the inequality $|\alpha | < 3 |f/(4\gamma )|^{2/3}$ reads as
\begin{eqnarray}
\label{b22}
|U^* - E| \leq \frac{3}{2} U_{12}^{1/3}\left (\frac{\Omega }{2}\right )^{2/3}
\equiv U_{12}^*
\, .
\end{eqnarray}
By the other words the characteristic interaction energy at the
intermediate region boundaries does not depend on $U_{12}$.
However, the positions of the linear turning points $|X^*|$
corresponding to the energies $U^* \pm U_{12}^*$ depend on the ratio
$U_{12}/U_{12}^*$. 
These points are located inside or outside
of the interval $[-a_0\gamma ^{-1/2} \, +a_0\gamma ^{-1/2}]$
at $U_{12}/U_{12}^* < 1$ and at 
$U_{12}/U_{12}^* > 1$, 
respectively. 
Accordingly for the both cases the matching conditions in the intermediate
energy region are different.
In the former case for the asymptotic matching region the potentials
can be reasonably approximated by parabola, and therefore we should work
with the Weber equations, and for the latter case the matching are performed
in the region where the potentials are linear ones, thus the equations are reduced
to the Airy equations. 

Let us investigate first the case
$U_{12}/U_{12}^* > 1$. Using Born-Oppenheimer approach of the section \ref{II},
the Schr\"odinger equations in the adiabatic representation with the accuracy up to $\gamma ^{-2}$
are decoupled for the wave functions $\Psi _\pm $
\begin{eqnarray}
\label{b23}
-\frac{d^2 \Psi _\pm }{d X^2} + \gamma ^2 (\alpha \pm \sqrt {v^2 + f^2 X^2}) \Psi _\pm = 0
\, .
\end{eqnarray}
The equations (\ref{b23}) at $|X| < v/f$ are reduced to the Weber equations with
the fundamental solutions 
$D_{-1/2 - q_1}(\pm \sqrt {2\gamma }X)$, 
and $D_{-1/2 +i q_2}(\pm \exp (-i\pi /4) \sqrt {2\gamma }X)$, where
\begin{eqnarray}
\label{b24}
q_1 = \gamma \frac{v + \alpha }{2} \, , \, q_2 = \gamma \frac{v - \alpha }{2}
\, ,
\end{eqnarray}
do not depend on the Massey parameter $\nu $.
Two real solutions of (\ref{b23}) correspond to the upper adiabatic potential
(classically forbidden region), and two complex solutions correspond to
the classically allowed  motion under the lower adiabatic potential. 

The argument of the Weber functions is $\propto X\sqrt \gamma $, and 
at the condition
$X < v/f$ their asymptotic expansions determine the interval where the matching 
should be done 
\begin{eqnarray}
\label{b25}
\gamma ^{-1/2}\left (\frac{U_{12}}{\Omega }\right )^{1/2} >  \gamma ^{-1/2}
\, .
\end{eqnarray}
Thus this inequality (\ref{b25}) can be fulfilled only at $U_{12}/U_{12}^* >1$,
when the intermediate region is sufficiently broad in comparison with $\Omega $.
In this case the exponents $q_1 \, , \, q_2$ (\ref{b24}) are large, 
and our aim is to find explicitely  the asymptotic expansions of the solutions in this case.
We will closely follow the method we borrowed from Olver paper \cite{OL59} (see
also his monograph \cite{OL74}),
which is in fact an expansion over small parameters $1/|q_i|$ (where $|q_i|$
are the exponents (\ref{b24}))
of the fundamental Weber solutions, and it
leads to the following asymptotic solution to Eq. (\ref{b23}) at $X > 0$
\begin{eqnarray}
\label{b26}
\Psi _+^-(X) \simeq Y_+^{-1/2} (X + Y_+)^{-q_1} \exp (- \gamma X Y_+)
\, , \,
\Psi _-^-(X) \simeq Y_-^{-1/2} (X + Y_-)^{iq_2} \exp (i \gamma X Y_-)
\, ,
\end{eqnarray}
where $Y_\pm = \sqrt {v \pm \alpha + X^2}$.
Using the known relation between the fundamental solutions of the Weber
equation \cite{EM53}, \cite{OL74}
$$
D_\mu (z) = \exp(-i\pi \mu )D_\mu (z) + 
\frac{\sqrt {2\pi }}{\Gamma (-\mu )}\exp \left (-i\pi \frac{\mu +1}{2}\right )D_{-\mu 
-1}(i z)
\, ,
$$
we can find two other (complimentary to (\ref{b26}) solutions
\begin{eqnarray}
\label{b27}
\Psi _+^+(X) = Y_+^{-1/2} \left (-\sin(\pi q_1)(X + Y_+)^{-q_1} \exp (- \gamma X Y_+)
+ \exp (-2 \chi _1) \frac{\sqrt {2\pi }}{\Gamma ((1/2) + q_1)}(X + Y_+)^{q_1}
\exp (\gamma X Y_+)\right )
\, ,
\end{eqnarray}
and
\begin{eqnarray} &&
\label{b28}
\Psi _-^+(X) = \\ &&
%\end{eqnarray}
%\begin{eqnarray}
\nonumber
Y_+^{-1/2} \left (-i\exp (-\pi q_2)(X + Y_-)^{iq_2} \exp (i\gamma X Y_-)
+ \exp (-2 \chi _2) \frac{\sqrt {2\pi }}{\Gamma ((1/2) - iq_2)}(X + Y_-)^{iq_2}
\exp (-i \gamma X Y_-)\right )
\, ,
\end{eqnarray}
where we introduce the notation
$$
\chi _1 = \frac{1}{2}\left (q_1 + \frac{1}{2}\right ) - \frac{q_1}{2} \ln \left
(q_1 + \frac{1}{2}\right ) \, , \,
\chi _2 = -\frac{1}{2}\left (iq_2 - \frac{1}{2}\right ) + \frac{i q_2}{2} \left
(-i \frac{\pi }{2} + \ln \left (q_2 + \frac{i}{2}\right )\right ) \, .
$$
Not surprisingly but it is worth to noting that these solutions (\ref{b26}) - (\ref{b28})
can be represented as a linear combination of the semiclassical solutions 
(\ref{b6}) $\Phi _\pm ^\pm $ with the coefficients
\begin{eqnarray}
\label{b29}
\cos 2 \theta _{(1 , 2)} = \frac{X}{\sqrt {v \pm \alpha + X^2}}
\, .
\end{eqnarray}
These energy dependent angles $\theta _{(1 , 2)}$ coincide with the adiabatic angles
(see (\ref{II11}) and (\ref{III12})) introduced above in sections \ref{II} and \ref{III} in
the level crossing point at $\alpha =0 $, and $ f|X| < v$, and the both
angles aquire only slightly different values over the whole intermediate region $|\alpha |
< v$.

Now we can find all needed connection matrices for these functions.
Although the calculation is straitforward it deserves some precaution 
(e.g. the $X$-dependent matrices have different functional form at the positive and negative
$X$). At $X > 0$ we get
\begin{eqnarray} && 
\label{b30}
\left( 
\begin{array}{c}
\Psi _-^- \\
\Psi _-^+ \\
\Psi _+^- \\
\Psi _+^+
\end{array}
\right ) =
\left [
\begin{array}{cc}
\cos \theta _2 & 0 \\
-i \exp(-\pi q_2) \cos \theta _2\, &\, \sqrt {2\pi }\exp(-2 \chi _2 )\cos \theta _2/\Gamma ((1/2) -
iq_2) \\
0 & 0 \\
0 & 0  
\end{array}
\right . \\ &&
%\end{equation}
%\begin{equation}
\nonumber
\left .
\begin{array}{cc}
0 & 0 \\
0 & 0 \\
\sin \theta _1 & 0 \\
-\sin (\pi q_1)\sin \theta _1 \, & \, \sqrt {2\pi }\exp (-2 \chi _1 )\sin \theta _1/\Gamma ((1/2) +
q_1) 
\end{array}
\right ]
\left (
\begin{array}{c}
\Phi _-^+ \\
\Phi _-^- \\
\Phi _+^- \\
\Phi _+^+
\end{array}
\right ) \, ,
\end{eqnarray}
and for $X < 0$ it reads as
\begin{eqnarray} && 
\label{b32}
\left( 
\begin{array}{c}
\Psi _-^- \\
\Psi _-^+ \\
\Psi _+^- \\
\Psi _+^+
\end{array}
\right ) =
\left [
\begin{array}{cc}
\sqrt {2\pi } \exp (-2\chi _2)\cos \theta _2/\Gamma ((1/2) - iq_2)\, &\, 
-i \exp(-\pi q_2)\cos \theta _2 \\
0 & \cos \theta _2 \\
0 & 0 \\
0 & 0
\end{array}
\right . \\ &&
%\end{equation}
%\begin{equation}
\nonumber
\left .
\begin{array}{cc}
0 & 0 \\
0 & 0 \\
\sin \theta _1 \sqrt {2\pi } \exp (-2 \chi _1)/\Gamma ((1/2) + q_1)\, &\, 
-\sin (\pi q_1)\sin \theta _1 \\
0 & \sin \theta _1 
\end{array}
\right ]
\left (
\begin{array}{c}
\Phi _-^- \\
\Phi _-^+ \\
\Phi _+^+ \\
\Phi _+^-
\end{array}
\right ) \, .
\end{eqnarray}
The product of the inverse to (\ref{b30}) matrix and the matrix (\ref{b32})
determines the connection matrix, we have sought for, to relate the semiclassical solutions in
the intermediate energy region 
(cf. (\ref{b20}) and (\ref{b21}) presenting the connection
matrices for
the tunneling and over-barrier energy regions). Performing this simple algebra
one ends up with
\begin{eqnarray} &&
\label{b33} 
U_{cross} =
\left [
\begin{array}{cc}
\sqrt {2\pi }\exp(-2 \chi _2 )/\Gamma ((1/2) - iq_2) & i\exp (-\pi q_2) \\
-i\exp (-\pi q_2) & 2 \exp (2 \chi _2) \Gamma ((1/2) - i q_2)\cosh (\pi q_2)  \\
0 & 0 \\
0 & 0 
\end{array}
\right . \\ &&
%\end{equation}
%\begin{equation}
\nonumber
\left .
\begin{array}{cc}
0 & 0 \\
0 & 0 \\
\sqrt {2\pi }\exp (-2 \chi _1 )/\Gamma ((1/2) + q_1 & \sin \pi q_1 \\
- \sin \pi q_1 &  \cos ^2 (\pi q_1)\Gamma ((1/2) + q_1) \exp (2 \chi _1)
\end{array}
\right ]
\end{eqnarray}
The matrix (\ref{b33}) has two $2 \times 2$ blocks structure, the same
as the connection matrices (\ref{b19}) and (\ref{b21}) for the tunneling and over-barrier
regions. However, unlike (\ref{b19}), (\ref{b21}) describing the
transitions between the diabatic states, the matrix (\ref{b33}) corresponds
to the transitions between the adiabatic states. Indeed, at a strong
level coupling ($U_{12} > U_{12}^*$) the eigenfunctions are close
to the adiabatic functions and only non-adiabatic perturbations induce the
transitions. Respectively, the off-diagonal matrix elements in (\ref{b33}),
having meaning of the probability to keep the same diabatic state after the transition,
are zero. The block with the real - valued matrix elements corresponds to the
minimum of the upper adiabatic potential, i.e. it is to the isolated
second order turning point, where \cite{BV02}
\begin{eqnarray}
\label{b34}
q_1 + \frac{1}{2} = \frac{U^* - E + U_{12}}{\Omega } + \frac{1}{2}
\, .
\end{eqnarray}
The complex - valued block is associated with the maximum of the
lower adiabatic potential, and analogously to (\ref{b34}) one
can find the following relation for the turning point
\begin{eqnarray}
\label{b35}
i q_2 + \frac{1}{2} = -i \frac{U^* - E + U_{12}}{\Omega } + \frac{1}{2}
\, .
\end{eqnarray}
In the case of weak level coupling, namely at $|U^* - E| < U_{12}^*$
and $U_{12} < U_{12}^*$, for the intermediate energy region, the
adiabatic potentials everywhere (except a small neighbourhood $|X| < v/f \to 0$
of the level crossing point) can be linearized, i.e. represented as
$\alpha \pm f|X|$, and the asymptotic solutions (\ref{b6}) are reduced
to a linear combination of the following functions
\begin{eqnarray}
\label{b36}
\Phi _+^\pm \propto (f|X|)^{-1/2}\exp (\pm \xi _+ sign X) \, ,
\,
\Phi _-^\pm \propto (f|X|)^{-1/2}\exp (\pm \xi _- sign X) \, ,
\, \xi _\pm = \frac{2}{3 f}(f |X| \pm \alpha )^{3/2}
\, .
\end{eqnarray}
Now all needed matrix elements can be calculated in the frame work of
the Landau perturbation theory \cite{LL65}, which can be formulated to avoid
divergency of the parameter $\nu $ at $\alpha \to 0$ in terms of
the dimensionless variables
$$
\tilde \alpha = 3\cdot 2^{-4/3} \frac{U^* - E}{U_{12}}
\, ; \, \tilde \nu = 3 \cdot 2^{-4/3} \frac{U_{12}}{U_{12}^*}
\, .
$$
The results of our analysis is shown in Fig. 6. The tunneling and the
over-barrier regions are separated from the intermediate energy region
by the lines $|U_{12}^* - E| = U_{12}^*$.
In own turn the intermediate region is also separated into two parts by the
line $\nu = \nu ^* = 0.325 $, where $\nu ^*$ is the value of the
Massey parameter $\nu $ at $U_{12}/U_{12}^* = 1$ and $|U^* - E| = U_{12}^*$.
In the $\nu < \nu ^*$ region the perturbation theory is the adequate tool
for the problem, and the transition matrix elements are proportional to
$U_{12}/U_{12}^*$. At $\nu > \nu ^*$ one can use the connection matrix
(\ref{b33}). To illustrate the accuracy of the approximations we have computed
the matrix element $M_{11}$. The results are shown in the Fig. 7.
Our computations demonstrate quite good precision, secured up to
two stable digits. The accuracy of the results on the boundaries
between the intermediate and over-barrier or tunneling regions is not
worse than $ 3 - 5 \% $, and can be even easily improved using
interpolation approaches.

\section{Scattering matrix.}
\label{VI}
LZ kind of phenomena can be considered as (and applied to) scattering processes.
Found in the previous section \ref{V} expressions for $4 \times 4$ connection
matrices can be used to calculate the scattering operator (or matrix) $\hat S$,
which converts an ingoing wave into an outgoing one. 

Let us consider first the over-barrier region for two linear
potentials crossing problem. For this case besides the crossing point
which we chose as $X=0$, there are two linear (first order) turning points
$X_0 = \pm |\alpha |/f $ (each turning point for each of the diabatic
potentials, we designate by $L$ and $R$). The scattering matrix which relates
the asymptotic solutions at $X \ll - X_0$ and at $X \gg X_0$ is the
product of the $4 \times 4$ connection matrix (\ref{b21}) calculated in the section
\ref{V} (subsection \ref{VA}) and known \cite{HE62} (see also \cite{BV02})
two semiclassical connection matrices describing wave function evolution from the turning point
$-X_0$ to the crossing point 0, and from this point to the turning point $+ X_0$, respectively.
Thus we end up with $2 \times 2$ matrix having the following block
matrix elements
\begin{eqnarray} &&
\label{d1} 
T_{11} = A_{if}
\left [
\begin{array}{cc}
\exp i(\phi - \phi _0) & 0 \\
0 & \exp -i(\phi - \phi _0)  
\end{array}
\right ]
\, ; \,
\\ &&
\nonumber
T_{12} = T_{21}^* = (1 - A_{if}^2)\exp (i\gamma W^*/2)
\left [
\begin{array}{cc}
i & -1/2 \\
-\exp -i\gamma W^* & (i/2)\exp -i\gamma W^*  
\end{array}
\right ] \\ &&
\nonumber 
T_{22} = A_{if}
\left [
\begin{array}{cc}
2\cos (\gamma W^* - (\phi - \phi _0)) & -\sin (\gamma W^* - (\phi - \phi _0)) \\
\sin (\gamma W^* - (\phi - \phi _0)) & (1/4) \cos (\gamma W^* - (\phi - \phi _0))  
\end{array}
\right ]
\, ,
\end{eqnarray}
where $A_{if} = (1 - \exp (-\pi \nu ))^{1/2}$ is the LZ amplitude of the transition
between the diabatic states, 
$\phi - \phi _0 = \tilde \chi $ (see (\ref{bd})),
and $W^*$ is the action between the linear turning points. 

The diagonal elements of (\ref{d1}), proportional to the transition amplitude
$A_{if}$, describe the propagating waves (i.e. the solutions
of the Schr\"oedinger equation in the lower adiabatic potential), and oscillating
blocks correspond to the solutions in the upper adiabatic potential. Off-diagonal
blocks proportional to the probability to keep unchanged the initial diabatic states,
describe the waves reflected from the linear turning points.
Interesting for physical applications reflection $R$ and transmission $T$ coefficients
can be found from (\ref{d1}) by a straightforward calculation, and the results
read as
\begin{eqnarray}
\label{d2} &&
R = -i (1 - A_{if}^2)[A_{if}^2 \exp (i\gamma W^* - 2 i (\phi - \phi _0))
+ \exp(- i \gamma W^*)]^{-1}  \, ;
\\ &&
\nonumber
T = 2 A_{if} \cos (\gamma W^* - (\phi - \phi _0))[A_{if}^2 \exp (i \gamma W^* - 2 i(\phi - \phi _0))
+ \exp (-i \gamma W^*)]^{-1}
\, .
\end{eqnarray}
The poles of the scattering matrix can be also easily found from (\ref{d1}),
and the corresponding resonance condition is
\begin{eqnarray}
\label{d3}
\cos (2(\gamma W^* - (\phi - \phi _0))) = - \left (1 - \frac{1}{2}
\exp (- 2 \pi \nu )\right )(1 - \exp (-2\pi \nu ))^{-1/2}
\, .
\end{eqnarray}
The action in the resonance points is complex-valued
\begin{eqnarray}
\label{d4}
Re (\gamma W^* - (\phi - \phi _0)) = \left (n + \frac{1}{2} \right )\pi
\, ; \,
Im (\gamma W^* - (\phi - \phi _0)) = - \frac{1}{2} \ln (1 - \exp (-2 \pi \nu ))
\, .
\end{eqnarray}
The poles of the scattering matrix are placed on the lower complex $E$ half 
plane at the vertical lines corresponding to conventional Bohr - Sommerfeld
quantization rules ($\gamma W^* = \pi (n + (1/2))$) for the upper
adiabatic potential. In the diabatic limit ($\nu \to 0$) the imaginary
part of the pole positions tends to infinity, and in
the adiabatic limit ($\nu \to \infty $) the poles move to the real axis.
Thus we see that the eigenstates of the upper adiabatic potential are
always quasistationary ones. The resonance widths are determined by the residues
of the scattering matrix elements at the poles, and it can be shown the resonance widths are
monotonically decreasing functions of $\nu $. In the Fig. 8 we show the energy
dependence of the transmission coefficient $T$. In the diabatic
limit $T \to 0$, and it is increased when $U_{12}$ is increased, and
in the over-barrier region there appear the resonances with the widths $\Gamma _n$
increasing with the energy increase, since in these conditions the Massey
parameter is decreased and $\Gamma _n \propto \exp (-2\pi \nu )$.

We illustrate the energy dependence of the transmitted wave phase in the Fig. 9.
In accord to the general scattering theory \cite{LL65}, there are $\pi $- jumps
of the phase at each quasi-discrete energy level of the upper adiabatic
potential. At $U_{12}/U^*_{12} < 1$, the resonance widths are of the order
of the inter-level spacings. The amplitudes of the decaying solutions
(localized in the well formed by the upper adiabatic potential)
are increased near the resonances, and this behavior is depicted in the
Fig. 10. One note of primary importance concerning the issue is that the
information about decaying solutions 
existing in the $4 \times 4$ connection matrix, e.g. (\ref{b21})
is lost when we use $2 \times 2$ scattering matrix (\ref{d1}).

Except for a slight natural modification the presented above results,
one can find the scattering matrix for the tunneling region  merely
by recapitulating the already derived expressions.
Thus instead of the matrix (\ref{d1}) we get
\begin{eqnarray} &&
\label{d5} 
T_{11} = 
\left [
\begin{array}{cc}
(1/4) M_{11}\exp (-\gamma W^*) + M_{22} \exp (\gamma W^*) & 
i((1/4) M_{11} \exp(-\gamma W^*) - M_{22} \exp(\gamma W^*)) \\
- i((1/4) M_{11} \exp (-\gamma W^*) - M_{22} \exp (\gamma W^*))  & (1/4) M_{11}\exp (-\gamma W^*)
 + M_{22} \exp (\gamma W^*)
\end{array}
\right ]
\, ; \,
\\ &&
\nonumber
T_{12} = T_{21}^* = \cos (\pi \nu )\exp (i\gamma W^*/2)
\left [
\begin{array}{cc}
i & -1/2\exp (-\gamma W^*)  \\
-1 & (i/2)\exp (-\gamma W^*) 
\end{array}
\right ] \\ &&
\nonumber 
T_{22} = 
\left [
\begin{array}{cc}
M_{11} & 0 \\
0 & M_{22}  
\end{array}
\right ]
\, ,
\end{eqnarray}
where $M_{11}$ and $M_{22}$ are the corresponding matrix elements from (\ref{b19}).

The same as above we compute the reflection and transmission
coefficients
\begin{eqnarray}
\label{d6} &&
R = -i \left (\exp (\gamma W^*) - \frac{1}{4} M_{11}^2 \exp (-\gamma W^*)\right )
\left (\exp (\gamma W^*) + \frac{1}{4} M_{11}^2 \exp (-\gamma W^*)\right )^{-1}
\, ;
\\ &&
\nonumber
T = M_{11} \left (\exp (\gamma W^*) + \frac{1}{4} M_{11}^2 \exp (-\gamma W^*)\right )^{-1}
\, .
\end{eqnarray}
In the intermediate energy region the only block matrix element $T_{11}$
requires the special calculations taking into account the contributions from
the complex turning points
\begin{eqnarray} &&
\label{d7} 
T_{11} = 
\left [
\begin{array}{cc}
\sqrt {2 \pi }\exp (-\pi q_2/2)/\Gamma ((1/2) - i q_2) & i \exp (-\pi q_2) \\ 
- i \exp (-\pi q_2) & 2 \Gamma ((1/2) - i q_2) \exp (-\pi q_2/2) ch (\pi q_2)/\sqrt {2\pi }
\end{array}
\right ]
\, ,
\end{eqnarray}
and all other matrix elements are the same as in the matrix (\ref{b32}).
Finally we find also in the intermediate energy region the reflection and the transmission coefficients
\begin{eqnarray}
\label{d8} &&
R = \frac{\exp (-\pi q_2)}{\sqrt {1 + \exp (-2\pi q_2)}}
\exp \left [-i (\phi - (\pi /2))\right ]\, ; \,
T = 
\frac{1}{\sqrt {1 + \exp (-2\pi q_2)}}
\exp (- i \phi )
\, ,
\end{eqnarray}
where $\phi = arg [\Gamma ((1/2) - i q_2)]$.
 
\section{Quantization rules for crossing diabatic potentials}
\label{VII}

In spite of the fact that instanton trajectories are rather simple objects,
and can be relatively easy found analytically, calculations of the
quantization rules within the instanton approach are rather involved
and intricate and require the knowledge of the scattering matrix
and all connection matrices, we have calculated in the previous sections.
In this section we apply this machinery to find the quantization
rules for the crossing diabatic potentials depicted in Fig. 11.
Depending on the Massey parameter the shown on the figure situations
exhaust all cases practically relevant for spectroscopy of non-rigid molecules
(symmetric or asymmetric double - well and decay potentials).

Within the instanton approach the quantization rule
can be formulated as a condition that the amplitudes of exponentially
increasing at $X > 0$, and $X< 0$, respectively, solutions $\Phi _L^+$, $\Phi _R^+$,
must be vanished. Taking into account that $W_L^* = W_R^*$ (the actions
in the corresponding wells of the lower adiabatic potential) and using
found above in the section \ref{IV} the connection matrix (\ref{b19}) ,
the quantization rule for this case is
\begin{eqnarray}
\label{e1} &&
\tan (\gamma W_L^*)  = \pm \frac{2}{p}\exp (\gamma W_B^*)
\, ,
\end{eqnarray}
where $W_B^*$ is the action in the barrier formed in the lower adiabatic potential,
and $p \equiv U_{11}$ is the corresponding matrix element of the
connection matrix (\ref{b19}).

Only the factor $1/p$ varying from 0 to 1 in the diabatic and in
the adiabatic limits, respectively, makes
this quantization condition (\ref{e1}) different
from the well known \cite{LL65} quantization rule for
the symmetric double-well potential. Correspondingly, the tunneling
splitting at finite values of the Massey parameter $\nu $ can be represented
as a product
\begin{eqnarray}
\label{e2} &&
\Delta _n = \Delta _n^0\, p(\nu )
\, ,
\end{eqnarray}
of the tunneling splitting $\Delta _n^0$ in the adiabatic
potential and the factor
\begin{eqnarray}
\label{e3} &&
p(\nu )= \frac{\sqrt {2\pi }}{\Gamma (\nu )} \gamma ^{\nu - (1/2)}\exp (-\nu )
\, ,
\end{eqnarray}
associated with the transition amplitudes between the diabatic
potentials in the crossing region.

It is particularly instructive to consider
(\ref{e1}) as the standard \cite{LL65} Bohr-Sommerfeld
quantization rule, where in the r.h.s. the both, geometrical
$\varphi _n$ and the tunneling $\chi _n$ phases are included additively.
In the adiabatic limit when $p(\nu ) \to 1$, we find that $\varphi _n \to 0$
and (\ref{e1}) is reduced to the quantization of the symmetric
double-well potential. In the diabatic limit $\varphi _n = - \chi _n$
and the geometric phase compensates the tunneling one. The physical
argument leading to this compensation may be easily rationalized as follows.
Indeed, at the reflection in the crossing point $ X= 0$, the trajectories
in the classically forbidden energy region are the same as those for
the tunneling region but with a phase shift $\pi $.

We focus now on the quantization rules for the over-barrier energy region.
Closely following the consideration performed above for the tunneling region
(replacing the connection matrix (\ref{b19}) by the matrix (\ref{b21}), and
making some other self-evident replacements) we end up after some tedious
algebra with the quantization rule
\begin{eqnarray}
\label{e4} &&
(1 -\exp (-2\pi \nu ))\cos (2\gamma W_L^* + (\phi - \phi _0))\cos (\gamma W^* -
(\phi - \phi _0)) + \exp (-2\pi \nu )\cos ^2 \left (\gamma W_L^* + \frac{\gamma W^*}{2}
\right ) = 0
\, ,
\end{eqnarray}
where $W^*$ is the action in the well formed by the upper adiabatic potential,
and $\phi - \phi _0 = \tilde \chi $ is determined according to (\ref{bd}).
From the Eq. (\ref{e4}) follows that the eigenstates are determined
by the parameter
\begin{eqnarray}
\label{e5} &&
B = \frac{\exp (-2\pi \nu )}{1 -\exp (-2\pi \nu )}
\, .
\end{eqnarray}
In the diabatic limit $\nu \to 0$, and therefore, $B \to 1/(2\pi \nu)$
in (\ref{e4}) the main contribution is due to the second term,
and it leads to a splitting of degenerate levels in the diabatic potentials.
Moreover since
\begin{eqnarray}
\label{e6} &&
\gamma \left (W_L^* + \frac{W^*}{2}\right ) = \pi \left (n + \frac{1}{2} \pm
2 \nu \sin \left [\gamma \left (W_L^* + \frac{W^*}{2}\right ) -\phi + \phi _0)\right ]
\right )  
\, ,
\end{eqnarray}
the splitting increases when the Massey parameter $\nu $ increases,
and it is an oscillating function of the interaction $U_{12}$.

In the adiabatic limit, when $\nu \to \infty $, $\phi - \phi _0 \to 0$, and,
therefore, from (\ref{e5}) $ B \simeq \exp (-2\pi \nu )$, the main contribution
to (\ref{e4}) comes from the first term which determines the quantization rule
for the upper one-well potential and for the lower double-well potential
in the over-barrier energy region, and in this limit the parameter $B$
plays a role of the tunneling transition matrix element. For $B$ smaller
than nearest level spacings for the lower and for the upper potentials, one can find
from (\ref{e4}) two sets of quantization rules leading to two
sets of independent energy levels
\begin{eqnarray}
\label{e7} &&
\gamma W^* = \pi \left (n_1 + \frac{1}{2}\right )
\, ; \,
2\gamma W_L^*  = \pi \left ( n_2 + \frac{1}{2}
\right ) 
\, .
\end{eqnarray}
Since the eigenstate energy level displacements depend on $U_{12}$
the resonances can occur at certain values of this parameter, where
the independent quantization rules (\ref{e7})
are not correct any more. The widths of these resonances
are proportional to $\exp (-2\pi \nu )$ and therefore
are strongly diminished upon the Massey parameter $\nu $ increase.
This behavior is easily understood, since in the limit the wave
functions of the excited states for the lower potential are delocalized,
and their amplitudes in the localization regions for the low-energy states
of the upper potential, are very small.

More tricky task is to derive the quantization rule in the intermediate
energy region. One has to use the connection matrix (\ref{b33}),
and to bear in mind the contributions from the imaginary
turning points. Nevertheless, finally the quantization rule can be written in
the simple and compact form as
\begin{eqnarray}
\label{e8} &&
\cos (2\gamma W_L^*) = - \exp (-\pi q_2)
\, ,
\end{eqnarray}
where $q_2 = \gamma (v - \alpha )/2$ is determined by the
relation (\ref{b24}).

It is useful to illustrate the essence of the given above general result
by simples (but yet non trivial) examples. First, let us consider two
identical parabolic potentials with their minima at $X = \pm 1$ and with
the coupling which does not depend on $X$. Since the symmetry,
the solutions of the Hamiltonian can be represented as symmetric
and antisymmetric combinations of the localized functions
\begin{eqnarray}
\label{e9} &&
\Psi ^\pm = \frac{1}{\sqrt 2}( \Phi _L \pm \Phi _R)
\, .
\end{eqnarray}
The functions are orthogonal, and, besides, two sets of the functions
$(\Psi _e^+ \, , \, \Psi _0^-)$, 
and $(\Psi _0^+ \, , \, \Psi _e^-)$ 
(where the subscripts $0$ and $e$ stand for the ground and for
the first excited states respectively) correspond to the two possible kinds of level crossings.

In Fig. 12 we depicted schematically the dependence of the
level positions on the coupling $U_{12}$. In the energy region $E \leq U^* + U_{12}$
where only there exist the discrete levels of the lower
adiabatic potentials, there are the pairs of the alternating parity
levels
$(\Psi _e^+ \, , \, \Psi _0^-)$, 
and $(\Psi _0^+ \, , \, \Psi _e^-)$. The tunneling splittings are increased   
monotonically since the Massey parameter $\nu $ is increased, and the
barrier is decreased with $U_{12}$. The same level and parity
classification is remained correct for the energy region above
the barrier of the lower adiabatic potential where the spectrum
becomes almost equidistant one. However, in the over-barrier region, the resonances
are occurred between the levels of the same parity, and this sequence
of the odd and of the even levels is broken, and level displacements
are not monotonic functions of $U_{12}$. Some of the levels of different
parities can be mutually crossed. For the upper adiabatic potential the level
sequence is opposite to this for the lower
adiabatic potential. Note that we checked the results of our
semiclassical approach and found remarkably good agreement
with the numerical quantum diagonalization. 

The second instructive example treats the one-well and linear diabatic 
potentials crossing. It leads to the lower adiabatic decay potential and
to the upper one-well adiabatic potential. The quantization rules in this case
correspond to the vanishing amplitudes for the exponentially increasing
solutions when $ X \to -\infty $, and besides one has to require
that no waves propagating from the region of infinite motion, i.e. at
$X > 1/2$. Performing the same as above procedure we find
that in the tunneling energy region, the eigenstates
are the roots of the following equation
\begin{eqnarray}
\label{e10} &&
\tan (\gamma W_L^*) = - i \frac{4}{p^2(\nu )}\exp (2 \gamma W_B^*)
\, ,
\end{eqnarray}
with the same as above notation.

To proceed further it is convenient to introduce the complex
action to describe the quasi - stationary states
\begin{eqnarray}
\label{e11} &&
\gamma W_L^* = \pi \left (\frac {E_n}{\Omega } - i \frac{\Gamma _n}{2\Omega }\right )
\, ,
\end{eqnarray}
where evidently $\Omega = \partial W_L/\partial E$ does depend on $E$.
From (\ref{e11}) the reel and imaginary parts of the quantized eigenstates
are
\begin{eqnarray}
\label{e12} &&
E_n = \Omega \left (n + \frac{1}{2} \right ) \, ; \,
\Gamma _n = p^2(\nu )\frac{\Omega }{2\pi }\exp (- 2\gamma W_B^*)
\, .
\end{eqnarray}
This relation (\ref{e12}) describes the non-adiabatic tunneling
decay of the quasi-stationary states of the lower adiabatic potential.
The same as we already got for the two parabolic potentials crossing
(\ref{e2}), here the tunneling and the adiabatic factors are entering
decay rate multiplicatively. Since the decay rate is proportional
to the square of the tunneling matrix element, $\Gamma _n \propto p^2(\nu )$
as it should be.

In the over-barrier energy region
the quantization rule is
\begin{eqnarray}
\label{e13} &&
(1 - \exp (-2\pi \nu ) \exp [-i (\gamma W_L^* + \phi + \phi _0)]\cos (\gamma W^* -\phi
+ \phi _0) + 
\\ &&
\nonumber
\exp (-2\pi \nu )\exp (-i \gamma W^*/2 ) \cos \left (\gamma W_L^* + \frac{\gamma
W^*}{2}\right ) = 0
\, ,
\end{eqnarray}
and the actions depend on the energy $E$ as
\begin{eqnarray}
\label{e14} &&
\gamma W_L^* = \pi \frac{E}{\Omega }
\, ; \,
\gamma W = \pi \left [ - \frac{U^* + U_{12}}{\Omega _1} + \frac{E}{\Omega _1}\right ]
\, ,
\end{eqnarray}
where $\Omega $ and $\Omega _1$ are $E$-dependent frequencies of the diabatic
and the upper adiabatic potentials.

In the diabatic limit the decay rate is proportional to the Massey parameter
and has a form
\begin{eqnarray}
\label{e15} &&
\Gamma _n \simeq \pi \nu \cos^2 (\gamma W -\phi + \phi _0)  
\, ,
\end{eqnarray}
and in the opposite, adiabatic, limit the decay rate is
\begin{eqnarray}
\label{e16} &&
\Gamma _n \simeq \exp (-2 \pi \nu )( 1 - \sin (2 \gamma W_L^* +\phi - \phi _0))  
\, .
\end{eqnarray}
In the both limits the decay rate is the oscillating function of $U_{12}$.
We illustrate the dependence $\Gamma (U_{12})$ for the 
crossing diabatic potentials $U_1
= (1 + X)^2/2$ and $U_2 = (1/2) - X$ in the Fig. 13.
Note that while the tunneling decay rate of the low-energy states is
increased monotonically with the Massey parameter $\nu $, the decay rate
of the highly excited states goes to zero in the both (diabatic and adiabatic)
limits. Besides there are certain characteristic values of $U_{12}$ when
the r.h.s. of (\ref{e15}) or (\ref{e16}) equal to zero and therefore $\Gamma _n =0$.

Last and more general example we consider in this section, describes
two non-symmetric potentials crossing at $X = 0$ point:
\begin{eqnarray}
\label{e17} &&
U_1 = \frac{1}{2}(1 + X)^2 \, ; \, U_2 = \frac{1}{2b}(X^2 - 2 b X + b)
\, .
\end{eqnarray}
In a certain sense it is the generic case, and when the parameter $b$ entering
the potential (\ref{e17}) is varied from 1 to $\infty $, we recover 
the two particular examples considered above,
and come from two identical parabolic potentials 
to the case one-well and linear diabatic 
potentials crossing. 
This kind of the potential $U_2$ was investigated recently by two
of the authors (V.B. and E.K) \cite{BK02}
aiming to study crossover behavior from coherent to incoherent
tunneling upon increase of the parameter $b$, the larger is this parameter
$b$, the larger will be the density of final states.
The criterion for coherent-incoherent crossover behavior found in 
\cite{BK02} based on comparison of the transition matrix elements and
the inter level spacings in the final state. The analogous
criterion should hold for LZ level crossing problem, however in the latter
case the tunneling transition matrix elements has to be multiplied by the small
adiabatic factor. Therefore the coherent - incoherent tunneling
crossover region moves to the more dense density of final states,
and the larger $U_{12}$ is the smaller will be the region for
incoherent tunneling.

Quite different situation occurs for highly excited states. In the diabatic limit, 
the transition
matrix element is increased with the Massey parameter $\nu $, and therefore
at a given $b$ value, the system moves to more incoherent behavior.
In the adiabatic limit, the transition matrix element is exponentially small,
and coherence of the inter-well transitions should be restored.
However, since the matrix elements are oscillating functions of $U_{12}$
for the intermediate range of this coupling ($U_{12}$) coherent - incoherent
tunneling rates are also non-monotonically varying functions.
To illustrate these unusual phenomena we show in Fig. 14 time dependence
of the survival probability $P$ for
the initially prepared localized in the left well state $n=0$.

\section{Conclusion.}
\label{VIII}

In this paper we have challenged again the very basic subject -
LZ problem. Currently there are about 100 publications per year
related to LZ problem. Clearly it is impossible to give a complete
analysis of what is being achieved in this field.
Our aim, therefore, was only to show some recent trends and our new results,
to help newcomers and specialists in finding cross-references between
the many physical phenomena related to LZ problem.
The problem was first addresses a long ago, 
and many already classical results are known now from the
textbooks \cite{LL65}, \cite{SL63}. 
Although exact quantum-mechanical calculations are 
still prohibitively difficult, many important results have been
obtained in the
frame work of the
WKB approach \cite{LL65} -\cite{MH96}. The accuracy of the modified WKB methods can be  
improved considerably, note for
example \cite{PK61} where the authors have included into the standard 
WKB method additionally a special type
of trajectories on the complex phase plane, 
along which the semiclassical motion is described by the Weber
functions. 
This method ascending to Landau \cite{LL65} is equivalent
to the appropriate choice of the integration path around
the turning point, and it appears to be quite accurate
for the tunneling and over-barrier regions, where the characteristic
fourth order polynomial (see (\ref{a15})) can be reduced to the second
order one (two pairs of roots are nearly degenerated). However, even in this
case there are some non-negligible corrections found in the papers \cite{ZN92} -
\cite{ZN94}. In the intermediate energy region, where all 4 roots are 
noticeably different, the method becomes invalid.
Besides 
the choice of these additional special trajectories (which one
has to include to improve the accuracy of the WKB method near the barrier top) depends on the detail form
of the potential far from the top, and therefore for each particular case the non-universal procedure
should be perform from the very beginning. 

We believe we are the first to explicitely addresses the question on
the behavior in the intermediate energy region. In all previous publications
this region was considered as a very narrow and insignificant one, or in
the best case the results were obtained by a simple interpolation
from the tunneling (with monotonic decay of the transition probability)
to the over-barrier (with oscillating behavior) regions.
The fact is that classical trajectories can be separated into two classes:
''localized'' and ''delocalized'' in the following sense.
If energy is close enough to the minimum or maximum of the potentials,
the trajectories could be called confined, since they are determined by the
universal features of the potentials in the vicinity of these
extremal points. Evidently it is not the case for the intermediate energy region.
In the present study we have found that, contrary to the common belief,
the instanton trajectory is a rather simple object and can be explicitely
computed even for the intermediate energy region.

Within the framework of the instanton approach we present
a full and unified description of $1D$ LZ problem, which is very often 
can be quite reasonable approximation for real systems.
Because different approaches have been proposed to study the LZ problem
we develop an
uniform and systematic procedure for handling the problem.
We reproduced all known results for tunneling and over-barrier regions,
and studied as well the intermediate energy region.
Specifically we applied our approach to the Born - Oppenheimer 
scheme, formulated the instanton method in the momentum space,
and presented all details of the LZ problem for two electronic states 
using also
the instanton description of the LZ problem in the coordinate space.
Neglecting higher order space
derivatives we find asymptotic solutions, and using adiabatic - diabatic
transformation we match the solutions in the intermediate region.
Based on these results we derived the complete scattering matrix for the LZ problem,
the quantization rules for crossing diabatic potentials.
Our results can be applied to several  models of level crossings 
which are relevant for the interpretation and description of
experimental data on spectroscopy of non-rigid molecules
and on other systems undergoing crossing and relaxation phenomena.

Note also, that in spite of the fairly long history of the LZ phenomena, the study is
still in an accelerating stage, and a number of questions remain to
be clarified (let us mentioned only few new features of the phenomena
attracted attention recently, like LZ interferometry for qubits \cite{SI01},
LZ theory for Bose - Einstein condensates \cite{YB02}, multi-particle and multi-level
LZ problems
\cite{PS02}, \cite{SI02}, \cite{DO00}, \cite{GC02}).
Much of the excitement arises from the possibility of discovering novel physics beyond say the
semiclassical paradigms discussed here. For example, we found in the sections \ref{II} and \ref{III}
that the wave functions of nuclei moving along the periodic orbits acquire geometrical
phases (the effect is analogous to the Aharonov - Bohm effect \cite{AB59}, but in our case
it has nothing to do with external magnetic fields and is related to
the non-adiabatic interactions). The relation between the both phenomena 
(the geometrical phases and the periodic orbits)
can be established using Lagrangian (instead of Hamiltonian) formulation of
the problem, which enables to take into account explicitely, using propagator technique,
\cite{MG72}, \cite{PS74}, \cite{MG75},
time dependence of the adiabatic process under consideration (see also, e.g.,
\cite{HL63},
\cite{KI85}). However, a proper handling of these aspects is beyond
the scope of our work. Further experimental and theoretical investigations
are required for revealing the detailed microscopic and macroscopic properties of 
different LZ systems.

In the fundamental problems of chemical dynamics and molecular spectroscopy, 
the transitions from the initial to final states can be treated as 
a certain motion along the
potential energy surfaces of the system under consideration.
These surfaces in own turn are usually determined within
the Born - Oppenheimer approximation (see section \ref{II}).
However, the approximation becomes inadequate for the excited vibrational
states, when their energies are of the order of electronic inter level energy
spacing or near the dissociation limit. In the both cases the non-adiabatic
transitions should be taken into account, and the most of the non-radiative
processes occur owing to this non-adiabaticity. The typical examples
investigated in the monography \cite{EL83}, are so-called pre-dissociation,
singlet-triplet or singlet-singlet conversion, and vibrational relaxation
phenomena. 

Slow atomic collisions provide other examples of the non-adiabatic
transitions between electronic states, where the time dependence of
the states is determined by the distance and by the relative velocity
of the colliding particles \cite{NU84}.
Some examples of the non-adiabatic
transitions relevant for semiconductor physics can be found in \cite{BD65},
for nuclear or elementary particle
physics in \cite{TO87} and for laser or non-linear optic physics in
\cite{JG78} - \cite{AK92}.
The latter topic of course is of interest in its own right but
also as an illustration of novel and fundamental quantum effects related to LZ model.
The off-diagonal electronic state interactions are arisen for this case from
the dipole forces. For relatively short laser pulses, it leads to the time dependent LZ
problem for two electronic states, detailed considered in our paper (see also
the laser optic formulation in \cite{JG78} - \cite{LG87}). The probability to
find the system in the upper state after a single resonant 
passage, can be computed in the frame work of
the LZ model. 
The latter point is related to one important aspect of the LZ problem, namely dissipative
and noisy environments. When external actions (say fields) driving LZ transitions
are reversed from large negative to large positive values, the dissipation
reduces tunneling, that is the system remains in the ground state,
or by other words, the thermal excitation from the ground state
to the excited one, suppresses such adiabatic transitions.
However, in the case of the field swept from the resonance point, 
the tunneling probability
becomes larger in the presence of the dissipation (see e.g., \cite{SK02}).
The increasing precision of experimental tests in the femtosecond laser pulse
range enables to excite well defined molecular states and to study their
evolution in time using the second probing laser beam \cite{GS92}.

\acknowledgements 
The research described in this publication was made possible in part by RFFR Grants. 
One of us (E.K.) is indebted to INTAS Grant (under No. 01-0105) for partial support.

\newpage

\centerline{Figure Captions.}
Fig. 1

Stokes (dashed) and anti-Stokes (solid) lines for a pair of close linear turning points
replaced by one second order turning point;
(a) - classically forbidden region;
(b) classically accessible region.

Fig. 2 

Adiabatic (3 , 4) and diabatic (1 , 2) potentials for LZ problem.

Fig. 3 

Stokes (dashed) and anti-Stokes (solid) lines in the vicinity of:
(a) conjugated bifurcation points $\pm i \tau _c$;
(b) diabatic potentials crossing point $X=0$.

Fig. 4

Stokes (dashed) and anti-Stokes (solid) lines for linear turning points corresponding
classically forbidden (a), intermediate (b), and accessible (c) energy regions of LZ problem.

Fig. 5

Relative placement of the adiabatic levels;
(a) $U_{12} > U_{12}^* \, ;$ (b) $ \, U_{12} < U_{12}^* $
 ,  $( U_{12}^* \equiv (3/2)(\hbar ^2 F^2/4 m)^{1/3}
).$ 

Fig. 6

$E$, $U_{12}$ phase diagram ($I$ - tunneling region, $II$ - over-barrier
region, and two intermediate energy regions $III$ and $III^\prime $
are separated by the line $\nu ^* =0.325$.

Fig. 7

Transition matrix element $M_{11}$ as
a function of $U_{12}/U_{12}^*$, computed at $\alpha = 0$:
on the
boundary between tunneling and intermediate energy regions (a);
at $ E = U^\# $ (b);
on the boundary between the intermediate and over-barrier regions (c);

lines $1 , 2 , 3 \, ,\, 1^\prime  , 2^\prime  , 3^\prime \, , \, 1^{\prime \prime } ,
2^{\prime \prime } , 3^{\prime \prime } $ computed for corresponding
energy regions using (\ref{b19}), (\ref{b22}), and (\ref{b34}), respectively.

Fig. 8

$T$ versus $E$ dependence for:
(a)  $U_{12} = U_{12}^*$ ;
(b)  $U_{12} =0.5 U_{12}^*$ ;
(c)  $U_{12} = 0.25 U_{12}^*$ ;

stars mark the region $III^\prime $ boundaries, thin lines show
results for the over-barrier and tunneling regions, bold lines for 
the intermediate energy region.

Fig. 9

Transmitted wave phase as a function of $E$ in the over-barrier region at
$U_{12} = U_{12}^*$.
 
Fig. 10

Amplitudes of the decaying solutions $\Phi _L^-$ at $X > 0$ versus $E$ for:
(1) $U_{12} = U_{12}^*$ ;
(2)  $U_{12} = 0.5 U_{12}^*$ ;
(3)  $U_{12} = 0.25 U_{12}^*$.

Fig. 11

The diabatic level crossing phenomena:
(a) crossing region;
(b) bound initial and decay final states;
(c) bound initial and final states.

Fig. 12

Level displacements versus $U_{12}$
for two diabatic crossing potentials
$(1 \pm X)^2/2$. Dashed lines show the intermediate
energy region; dotted - dashed lines show displacements for the top
and for the bottom of the adiabatic potentials. $k$, $n$, and $n^\prime $
are quantum numbers for the diabatic, and lower and upper adiabatic potentials.

Fig. 13

$\Gamma _n$ versus $U_{12}$ for the quasi stationary states
at the diabatic potentials $(1 + X)^2/2$ and $(1/2) - X$ crossing;
(a) 1 - 4 are the level energies 0.042 , 0.125 , 0.208 , and 0.292 
for the lower adiabatic potential;
(b) $1^\prime - 3^\prime $ are the level energies 0.625 ; 0.708 ; 0792
for the upper adiabatic potential. 

Fig. 14

Survival probability for the localized $n=0$ state; 
(a) $b = 1500$ , dashed lines $U_{12} = 0.15$ ; solid lines 
$U_{12} = 0.21$;
(b) $b = 1500$, dashed lines $U_{12} = 0.28$ ; solid lines  
$U_{12} = 0.21$.

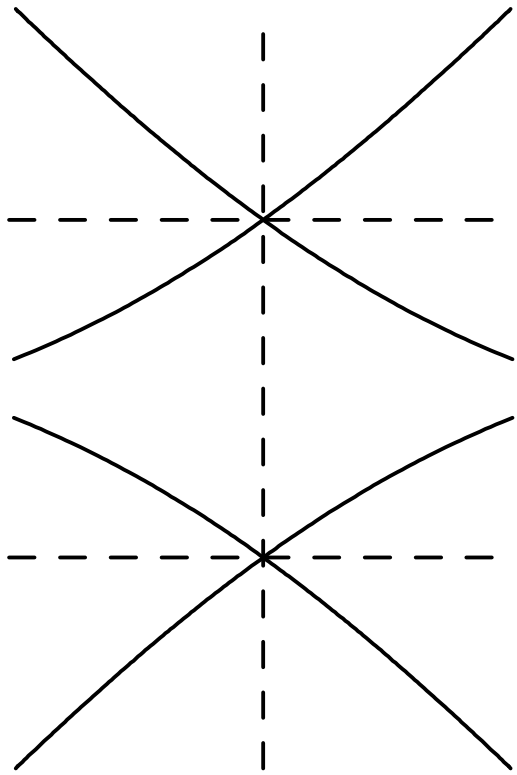
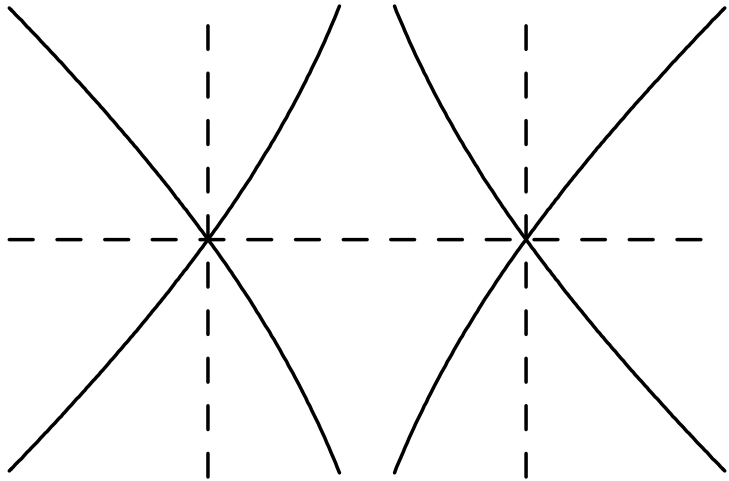
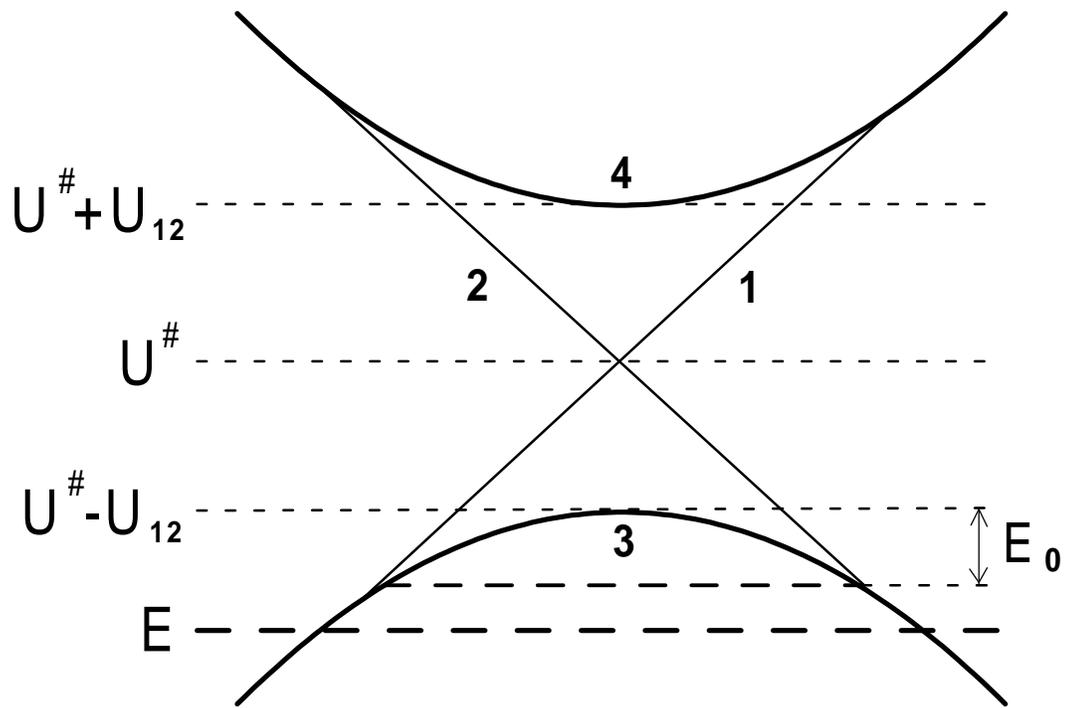
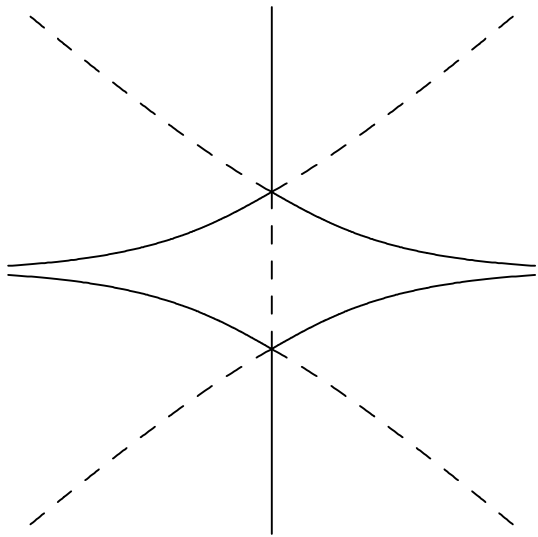
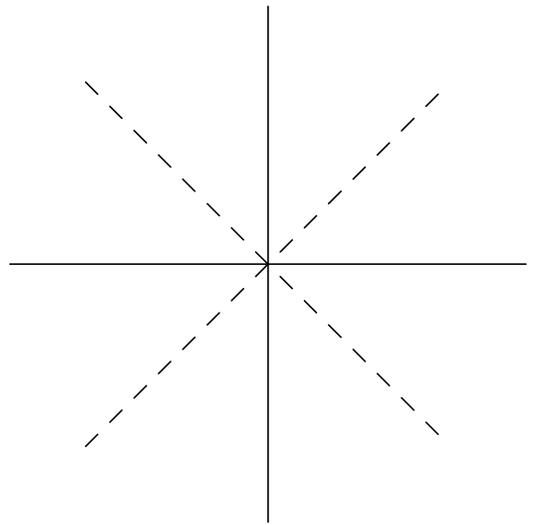
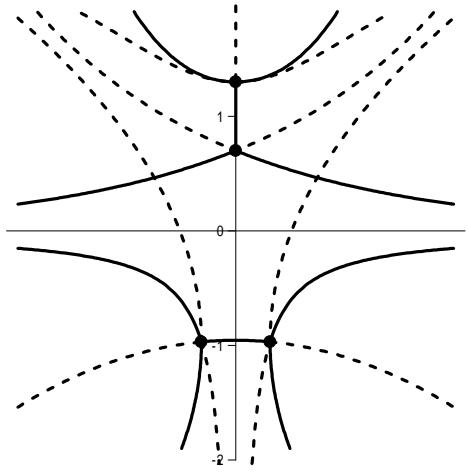
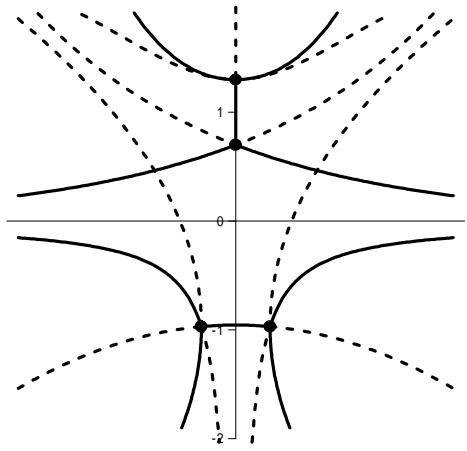
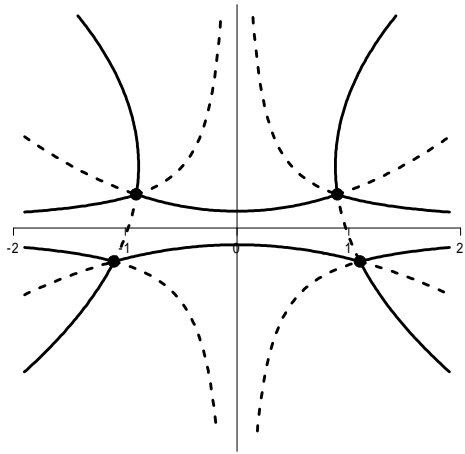
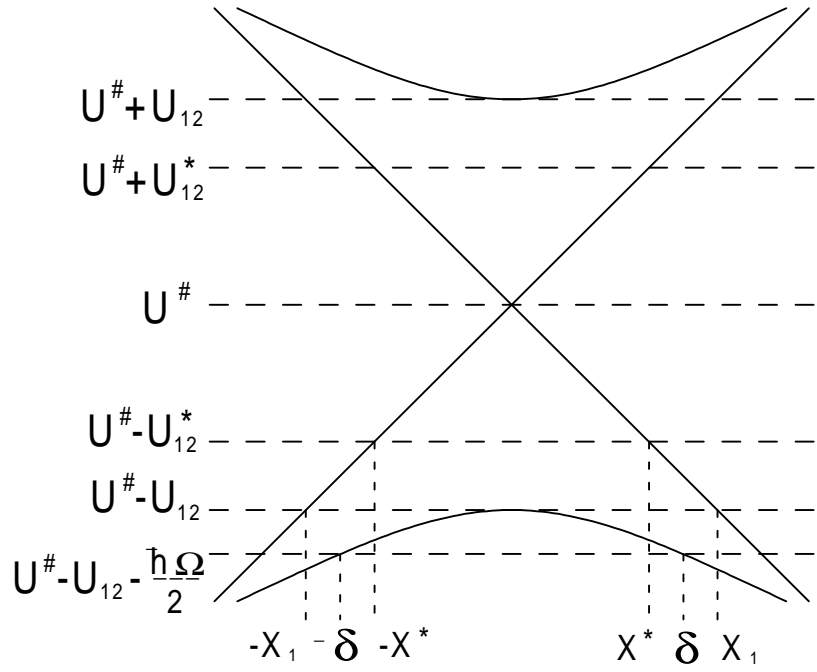
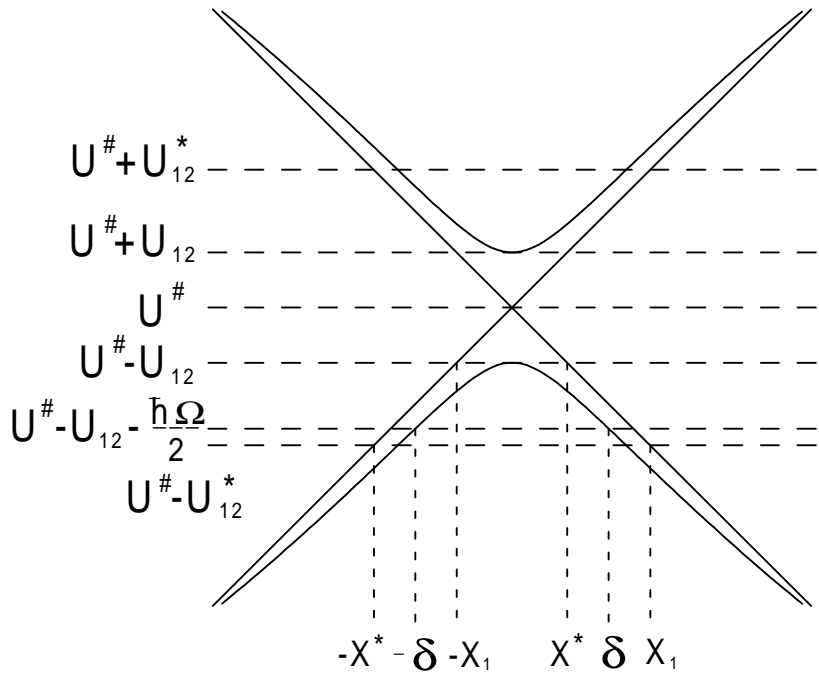
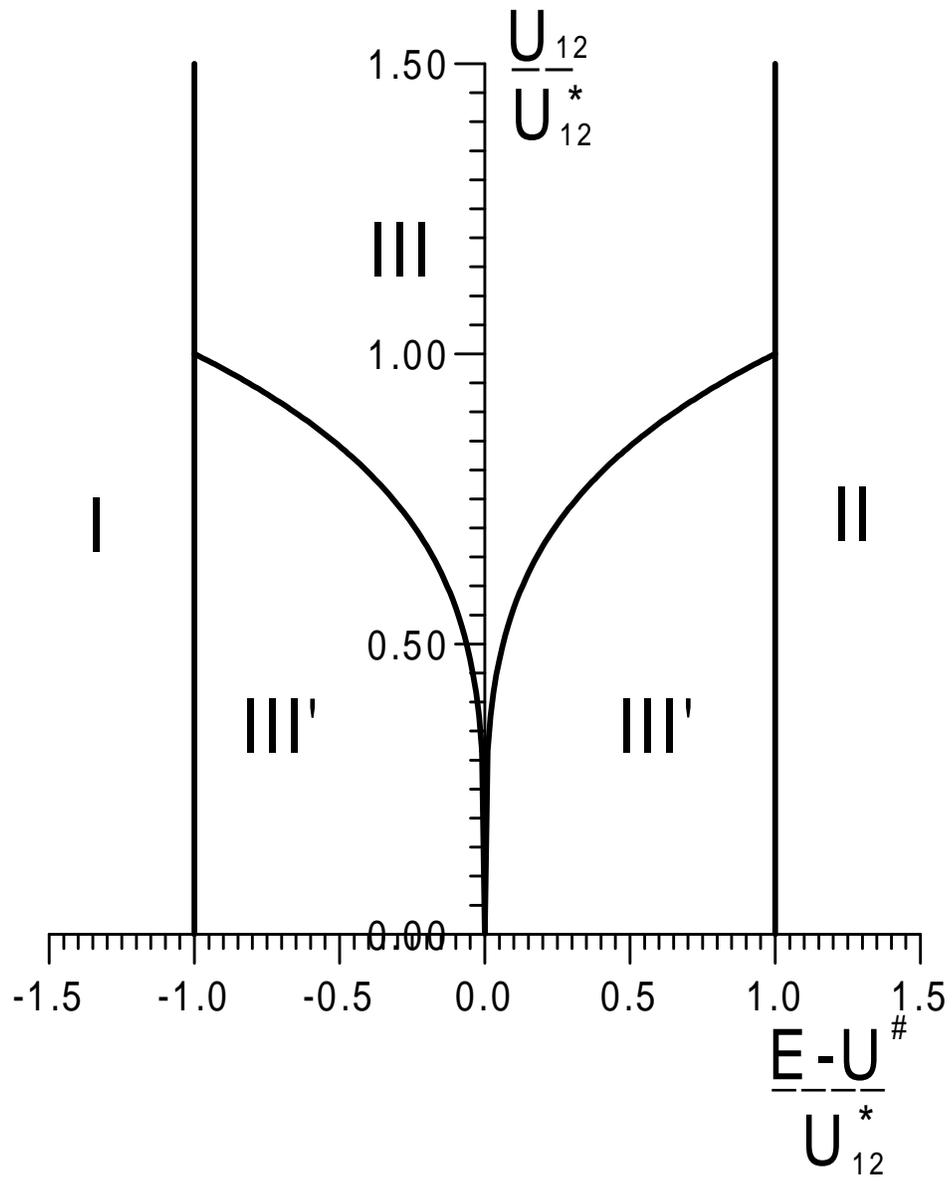
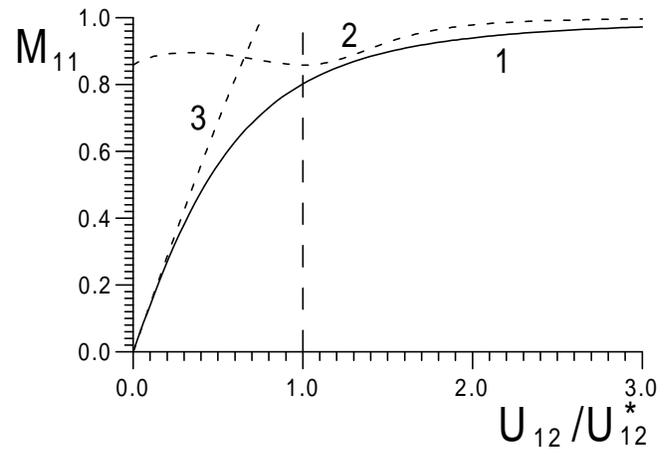
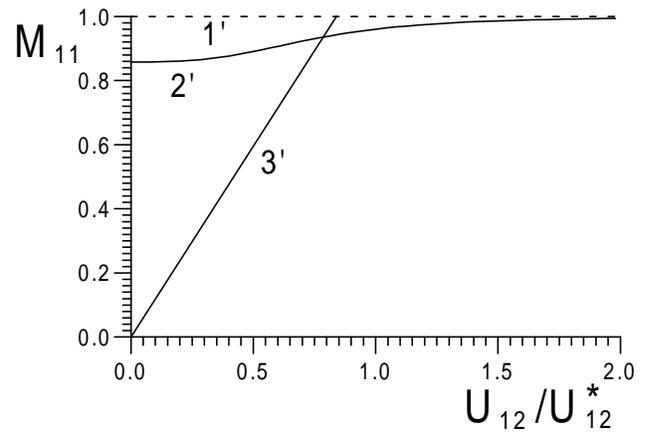
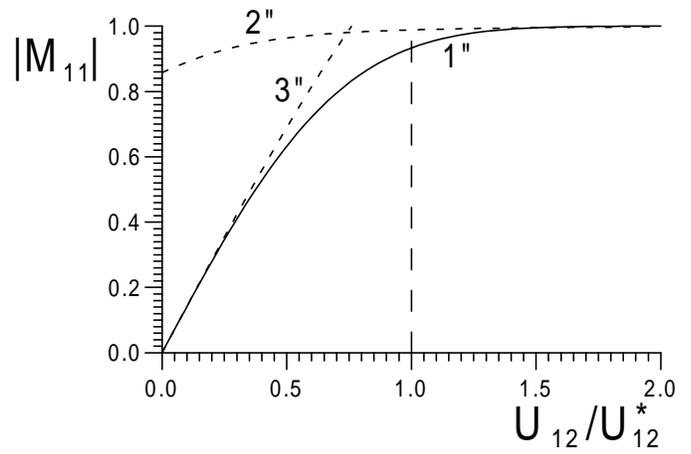
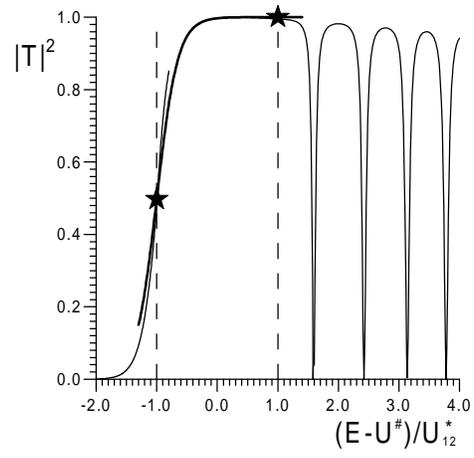
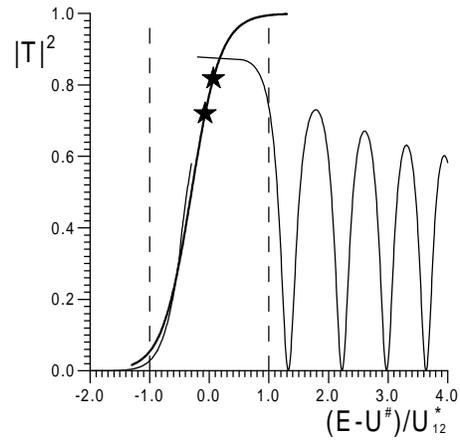
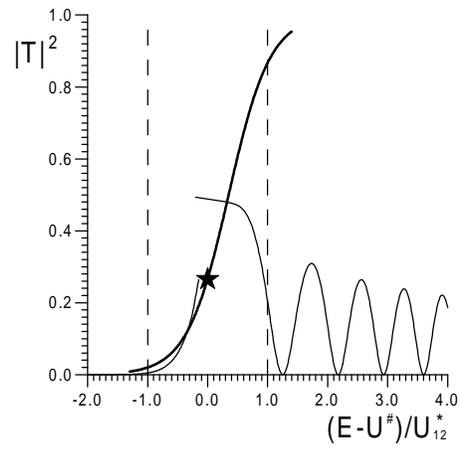
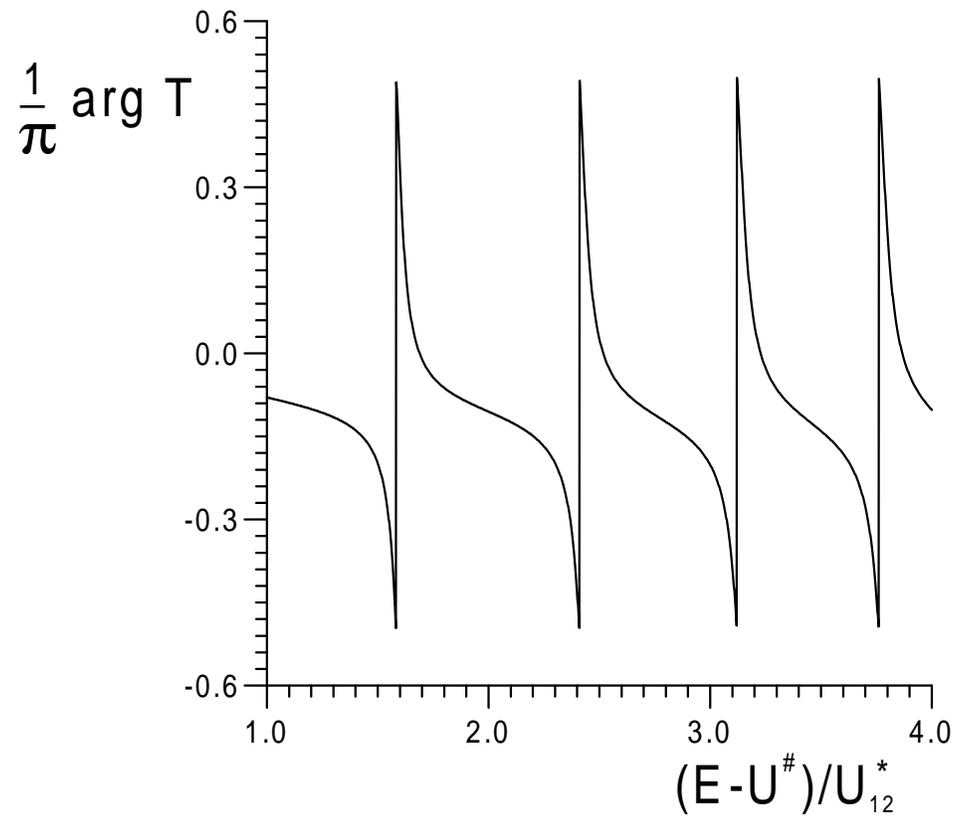
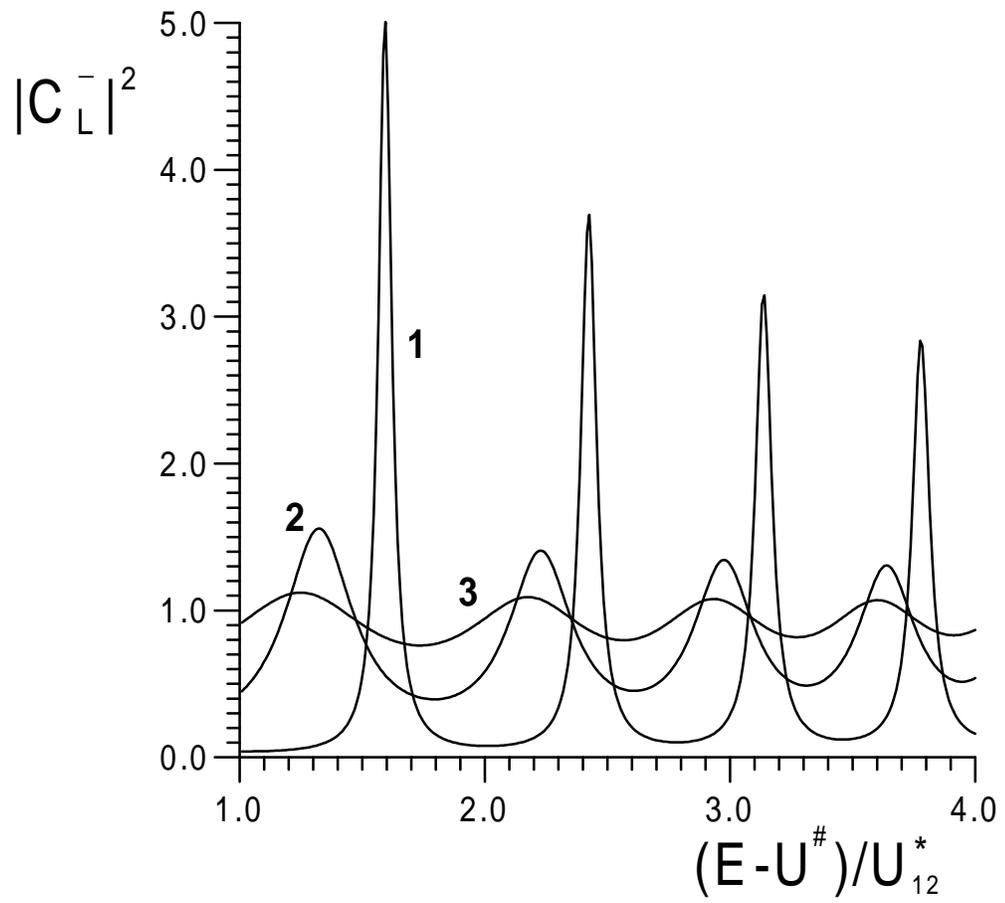
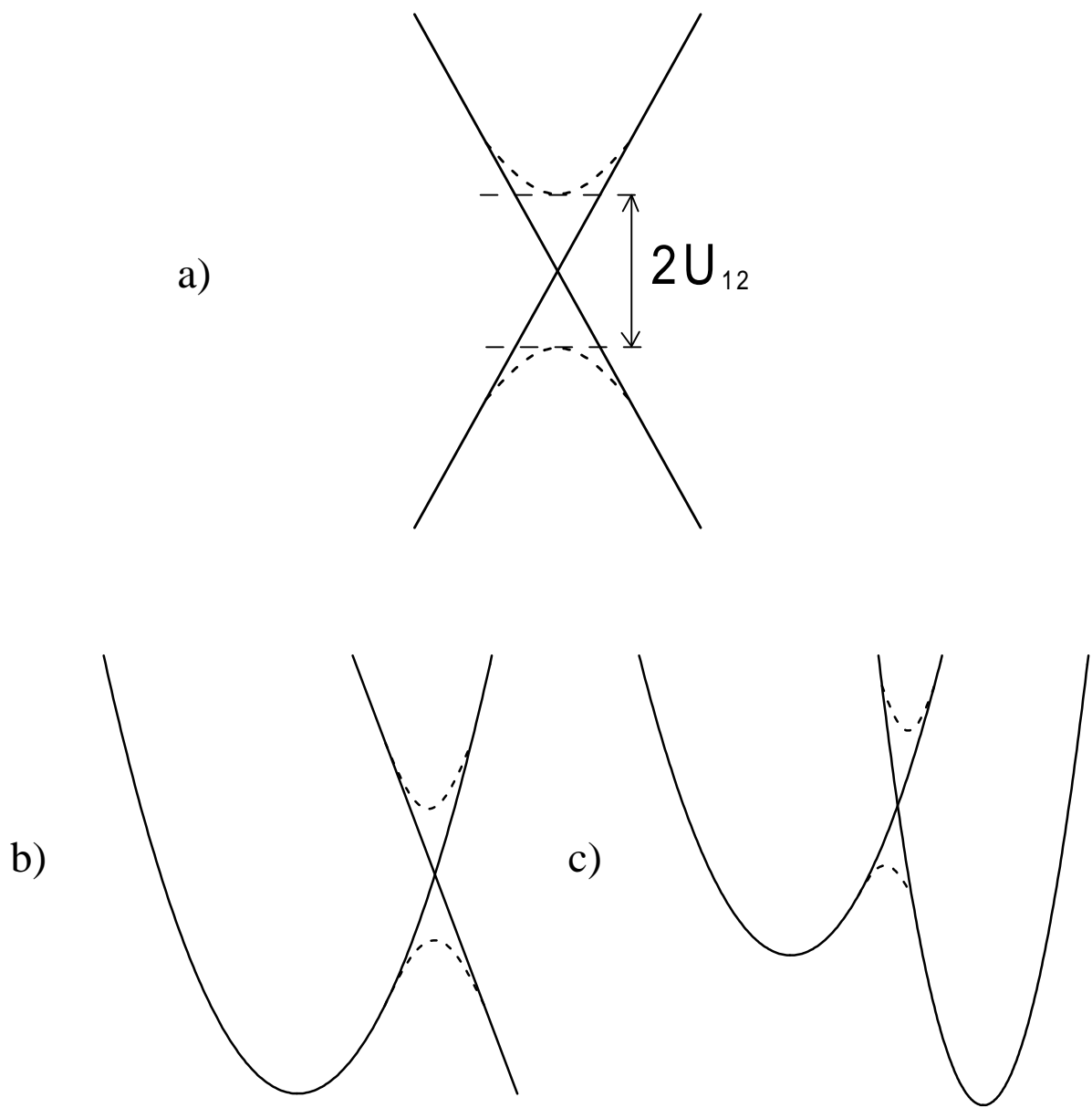
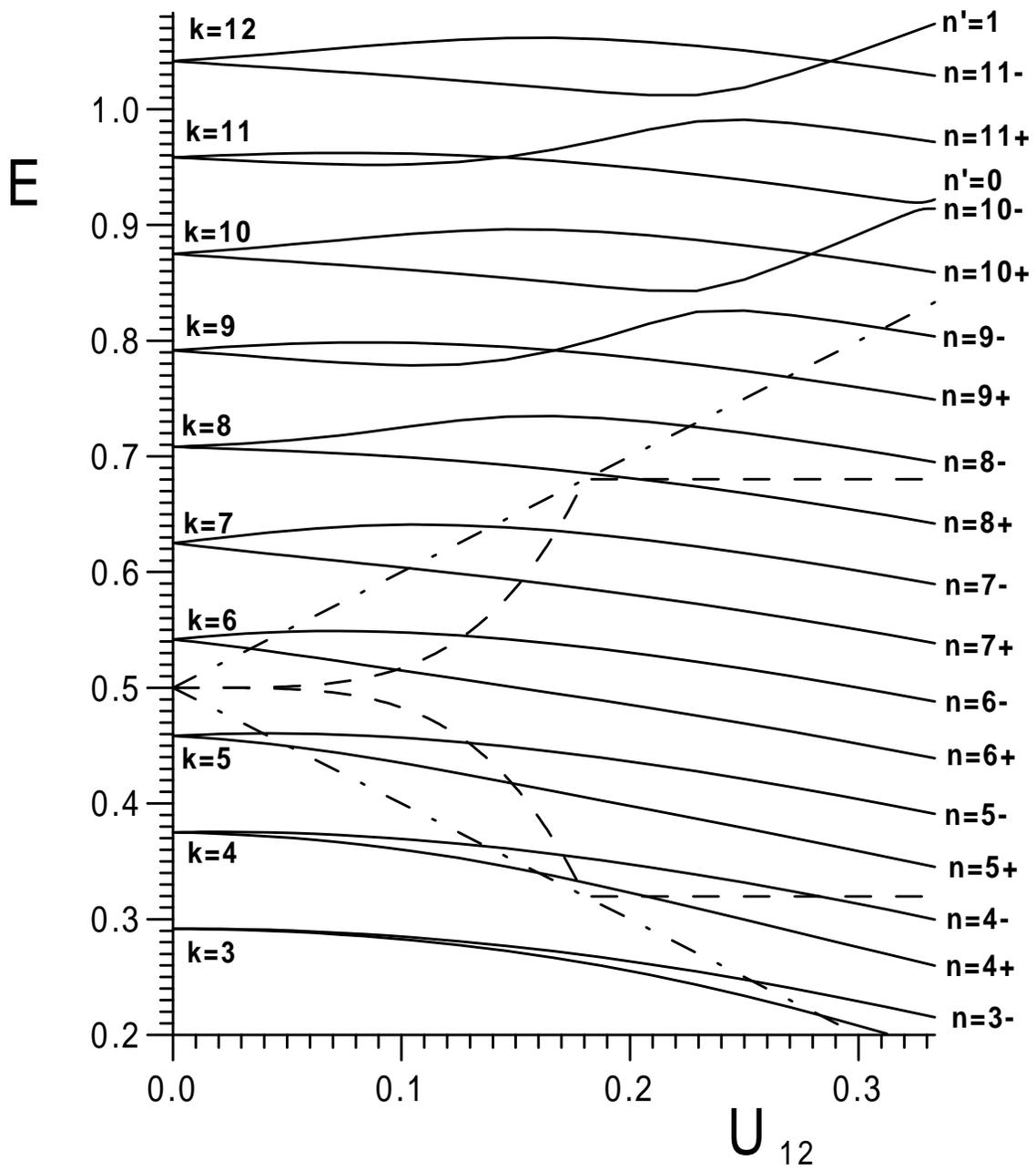
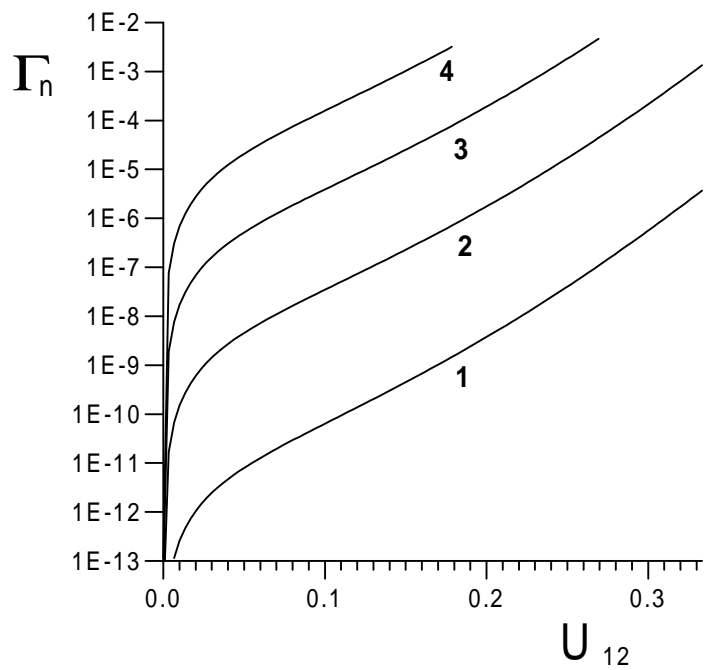
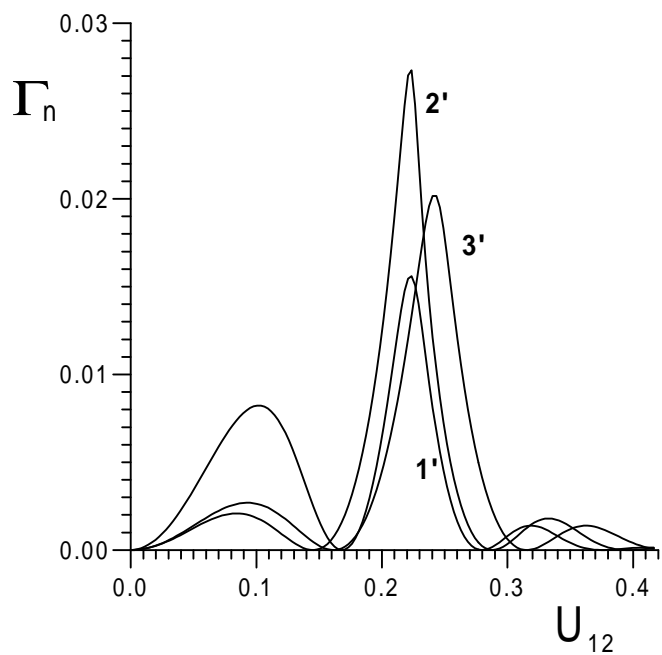
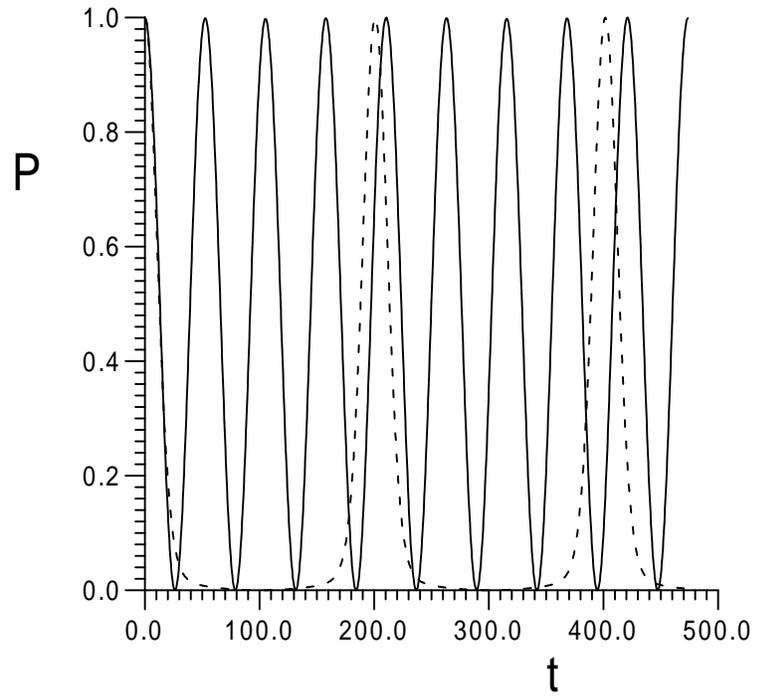
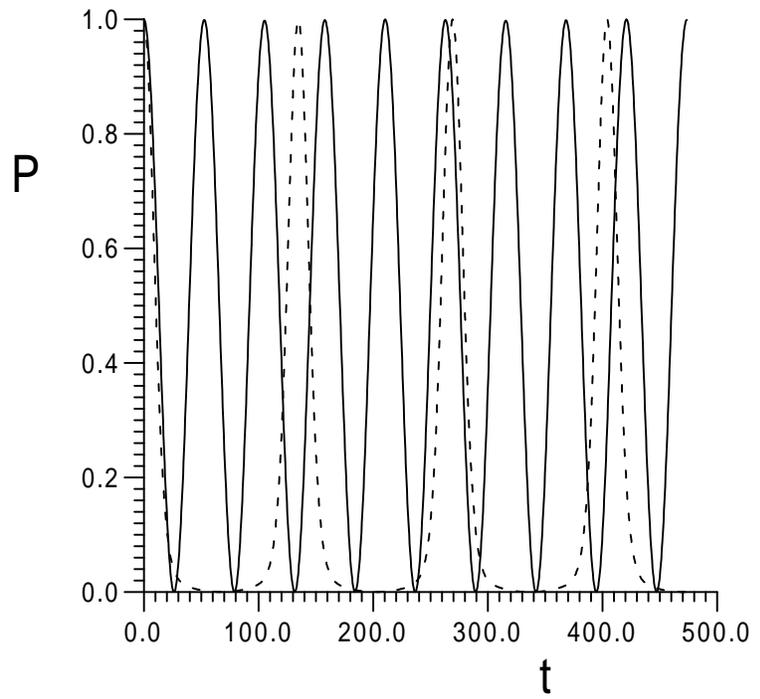
\end{document}